# Developing a Framework for Heterotopias as Discursive Playgrounds:

# A Comparative Analysis of Non-Immersive and Immersive Technologies

Elif Hilal Korkut[1]  and Elif Surer[1, *]

[1] *Department of Modeling and Simulation, Graduate School of Informatics, Middle East Technical University, 06800 Ankara, Turkey*

**\*Corresponding Author:**
Assoc. Prof. Elif Surer
Department of Modeling and Simulation
Graduate School of Informatics
Middle East Technical University
06800 Ankara, Turkey
elifs@metu.edu.tr

**Elif Hilal KORKUT** *(elif.korkut@metu.edu.tr) is a Master's student in the Multimedia Informatics program at Middle East Technical University with a Bachelor's degree in Architecture. She is interested in generative algorithms, computational design, exhibition and display design, vision and visuality, and game technologies.*

**Elif SURER** *(elifs@metu.edu.tr) is currently working as an Associate Professor with the Graduate School of Informatics' Multimedia Informatics program at Middle East Technical University. Her research interests include serious games, virtual/augmented reality, and reinforcement learning.*

# Developing a Framework for Heterotopias as Discursive Playgrounds: A Comparative Analysis of Non-Immersive and Immersive Technologies

## Abstract

The discursive space represents the reordering of knowledge gained through accumulation. In the digital age, multimedia has become the language of information, and the space for archival practices is provided by non-immersive technologies, resulting in the disappearance of several layers from discursive activities. Heterotopias are unique, multilayered epistemic contexts that connect other systems through the exchange of information. This paper describes a process to create a framework for Virtual Reality, Mixed Reality, and personal computer environments based on heterotopias to provide absent layers. This study provides virtual museum space as an informational terrain that contains a "world within worlds" and presents place production as a layer of heterotopia and the subject of discourse. Automation for the individual multimedia content is provided via various sorting and grouping algorithms, and procedural content generation algorithms such as Binary Space Partitioning, Cellular Automata, Growth Algorithm, and Procedural Room Generation. Versions of the framework were comparatively evaluated through a user study involving 30 participants, considering factors such as usability, technology acceptance, and presence. The results of the study show that the framework can serve diverse contexts to construct multilayered digital habitats and is flexible for integration into professional and daily life practices.

## 1. Introduction

Every culture, field, and individual creates different ways of communicating and constructing knowledge by selecting certain elements, excluding others, and rearranging them in a particular order. The formation and constraints can vary, but the meaning and knowledge production occur through narration and interpretations of these units (Maciag, 2018). The increasing availability and accessibility of digital information, made possible through processes like digitalization and digitization, has changed discursive practices. Actors have been narrating emerged "formless chaos of knowledge" composed of multimedia via technological systems to produce, consume, and preserve information (Manovich, 2001). While technology provided accessibility and usability to users, extensive use of systems facilitated by the widespread adoption of technology and the increasing reliance on personal computers and mobile phones led to a transformation, elimination, and dissociation of several layers in daily life and professional practices. Odom et al. (2014) describe this situation as "placelessness" and "formlessness." In the absence of certain layers, the user constantly maps the physical interactions and customizations to "virtual possessions" in the digital environment.

Being a multilayered discursive space, the missing layers of this new territory can be explored with recourse to Foucault's heterotopias, "areas of resistance that intensify knowledge" (Foucault, 2008). In this study, we investigated the layers essential for comprehensive production, representation, and interpretation processes for professional and personal "archival practices" through the concept of heterotopia. Foucault provides several strategies to construct discourses: "operating table," "surfaces of emergence," "authorities of delimitation," and "grids of specification" (Foucault, 2019). Based on layers of heterotopias and their potential to integrate strategies of discourse, we provided a framework that can support users in creating their digital habitats via immersive technologies incorporating individual archives.

Resistance of heterotopias occurs through adaptations and changes in function resulting from dynamic power relations, which is an essential layer of heterotopias. The individual is regarded as a "focal point of resistance" in modernity, with the ability to shape and manipulate experiences and products in conjunction with the systems to which they are linked (Thompson, 2003). Together with the system, they become "authorities of delimitation." The developed framework offers the dynamism of heterotopias that can be altered to fit the needs of users, with the ability to import various formats to compose a content layer and modular structure based on object-oriented programming. It empowers individuals to create experiences based on their content in an immersive environment where they become the authority.

Heterotopias provide alternative viewpoints of the world and opportunities for people to explore different ways of thinking by juxtaposing and combining many spaces into one site; they problematize received knowledge from the "surfaces of emergence." The framework functions as an "operating table," it expands the space of heterotopias and enables users to interact with the elements of multimedia language in changeable settings. Heterotopias are dynamic physical or virtual spaces that exist outside of the normal ordering of time and space. They offer spatial layers to reorder information (Foucault, 2008). The presented framework offers "grids of specification," forming "liquid architectures" where users can create places according to their needs (Novak, 1992). It includes procedural generation algorithms (PCG) that provide different levels of autonomy while offering several architectural elements. This way, the framework provides another layer that can be interpreted as the meaning of the space that contains complex relations. This layer includes how the space is narrated, represented, or understood by different groups of people and the contexts that give it meaning (Lehtinen, 2022).

Through embodiment, mixed and virtual reality technologies have the potential to create a spatial layer of heterotopias by creating immersive, interactive digital environments that can mimic the feeling of being in a physical place or blend the physical space with the virtual and provide a sense of place and connection to their surroundings (Schultze, 2010). Virtual Reality (VR) and Mixed Reality (MR) technologies are extensively studied to facilitate places and knowledge production by allowing users to engage in activities and simulations of cultural heritage (Cecotti, 2022; Gonizzi Barsanti et al., 2015), construction management (Hepp and Hasebrink, 2018; Safikhani et al., 2022), and architecture (Akin et al., 2020; Prabhakaran et al., 2022). While these studies provide the existence of spatial layers, they were developed to target specific fields. Therefore, they do not offer comprehensiveness in terms of multimedia and provide an operating table for only targeted fields which eliminates the possibility of using them in other contexts.

In this study, virtual museum space was specifically chosen as an overarching term that can contain various types of entities holding both place and media representation and reproduction layers. Exhibitions inside a virtual museum create a "world within worlds" through narration constructed with places. Foucault (2008) already defined museums as heterotopia; however, not all museums are necessarily discursive heterotopias. The current applications show that large corpora consisting of virtual museums provide abstract or exact copies of a real museum with isolated interactions based on predefined paths (Cecotti, 2022; Kersten et al., 2017). There is an increase in the number of studies improved by individualization being incorporated in a static, designed, built environment (Foo et al., 2009; Komianos and Oikonomou, 2018). Virtual museum applications offer a wide range of multimedia elements; however, they provide static functions, and personal content has been limited to images. A system that provides the technical capability of integrating diverse personal media such as text, audio, and 3D graphics is not available. Also, the elements are provided in digital "emplacements." They do not offer the adaptable place production required for reordering individual content. Therefore, they mostly do not provide operating table functions, including the resistance presented by individuals.

Building on these insights, we developed a flexible framework that can provide the construction of heterotopias for the broader audience and disciplines, which makes the framework usable for further research and diverse practices. The spatial layers of heterotopia creation are provided via architectural elements open for manipulation and texture, lighting, and scale options and supported via integration of PCG algorithms, especially for novice users. Additionally, by injecting user preferences, we provided different automation levels. Content layers and archival practices are provided via several algorithms that can import, store, and exhibit various data formats and allow analysis and grouping of personal content. In conjunction with the ability to hold various layers, the developed framework can provide a playground based on heterotopias. Throughout the design process, we followed meta-design principles, which helped us to create a flexible and modular framework that can shape via needs of users to serve diverse contexts.

This study compares experiences across three platforms (PC, VR, MR) to understand the effects of different technologies, which can serve as further insights to increase the integration of immersive technologies. We conducted a comparative user study with the participation of 30 respondents. After exposure to the system, participants completed standardized presence, technology acceptance, usability questionnaires, and system-specific surveys. Later, we conducted semi-constructed interviews to gain better insights and interpret the objective and subjective data gathered in the previous steps. Results indicate that we managed to provide standard levels of usability and acceptance across technologies, and the framework is suitable for diverse contexts.

While VR and MR technologies can offer absent layers of 2D interfaces, from the meta-design perspective (Fischer et al., 2004; Lee et al., 2020), when compared to PC-based tools, immersive technologies are still waiting for more pervasive use, which requires new human-computer interaction (HCI) approaches to integrate those technologies into professional and personal life on a daily basis, to create a comprehensive dialogue between system, user, content, and place.

To derive design features that can increase the integration of immersive technologies into different contexts, this study aims to provide answers to the following questions:

R1. In the creation process of heterotopias, which medium provides better experiences in terms of usability, presence, and technology acceptance?

R2. According to which dimensions do technology acceptance, usability, and presence of versions present differences?

R3. Which medium is more applicable for the interaction with different asset types?

R4. How do different mediums change the design approach of users?

R5. How do scale, point of view, lighting, and material aspects affect users' experiences and tasks?

R6. What are the effects of the autonomy level of procedural generation algorithms on users' preferences?

In what follows, we review works related to immersive technologies, virtual museums, and procedural generation techniques, and to facilitate readability and to preface our user analysis, we briefly outline the conceptual terms that emerged through technology. Then, we describe each aspect of the proposed application and the rationale behind the design decisions. In the next part, we explain our evaluation methods and procedures. Then, we provide the results of the user tests about related concepts from various fields. We discuss the results while reflecting on the extended concepts and terms that emerged through user tests. Finally, to conclude, we summarize our study and add insights into how further details might be searched.

## 2. Literature Review

### 2.1. Discourse and Heterotopias

Michel Foucault was a French philosopher who studied the ways in which power and knowledge shape society. He identified certain places, called heterotopias, that exist within a culture and serve as a sort of "realized utopia" where other places within the culture are represented, disputed, and reversed. Heterotopias are characterized by their ability to juxtapose and compare various spaces and concepts and can manifest in various architectural forms, functions, and sizes (Foucault, 2008).

According to Michel Foucault, discourse, or systems of thought and representation, plays an important role in shaping our understanding of the world. He identified several strategies for constructing discourses: "operating table," "surfaces of emergence," "authorities of delimitation," and "grids of specification." The operating table represents the space in which knowledge is constituted serving as a means of classifying and ordering things, allowing us to make sense of them. Surfaces of emergence are the foundation for the development of new discourses and the creation of new subjects, while an authority of delimitation is a set of rules that determines how objects can become subjects within a discourse. The grids of specification are a taxonomy of concepts used to order objects into a hierarchy within a discourse. By understanding these strategies, Foucault explained how discourses shape our understanding of the world and our actions within it (Foucault, 2005).

Heterotopias are distinctive spaces that can contain the processes of discourse. Michel Foucault identified various characteristics of these spaces and provided examples to help illustrate and understand these features. According to Michel Foucault, all cultures produce heterotopias of deviation and crisis such as boarding schools and prisons. Depending on the culture and time period, these spaces may serve a variety of purposes within a society, and their features and functions evolve over time, which provides resistance. He argued that resistance is not separate from power relationships but rather is an integral part of them. Power is not a fixed entity but is constantly negotiated and contested through interactions, leading to changes in the functions of heterotopias (Lehtinen and Brunila, 2021). Heterotopias are spaces that can bring together multiple incompatible spaces and elements in a single location, disrupting the way we understand the world. Foucault also investigated the "order of discourse" as the organization of the rules and systems that govern knowledge production (Foucault, 2005). Heterotopias offer alternative ways of organizing information and the layout of a space can affect the way knowledge is shared and understood, providing "grids of specification." The characteristics and purpose of these spaces influence the meanings and perceptions that can be experienced through embodiment.

Heterotopias are spaces that are connected to certain periods in time. They can either permanently store time, like archives or libraries, or they can provide a brief window into a specific time, like exhibitions or festivals. In modern society, the individual becomes the "focal point of resistance," and archives, libraries, and museums often represent personal choice and "self-formation," with the individual acting as a point of resistance to the shaping power of the larger system they exist within. These individuals become "authorities of delimitation," shaping their own products within the constraints of the system (Foucault, 1972).

Heterotopias are places where access is controlled, meaning that they are not freely accessible like public spaces. This control can be achieved through technological means, such as the use of devices that have opening and closing systems or personal accounts. Heterotopias are sites that have a particular connection to the environment around them. They may either expose the essence of other real spaces or create an organized space that is distinct from the disorder of other places. In the context of discourse, this relationship creates "surfaces of emergence," where new ideas, arguments, or perspectives arise or come to the forefront of the discourse (Foucault, 1972).

## 2.2. Heterotopias and Digital Playgrounds

Rousseaux and Thouvenin (2009) explored Informed Virtual Sites (IVS) through Michel Foucault's heterotopias by superimposing digital and physical spaces. They also named these heterogeneous places heterovirtopias as an extension of Foucault's terminology. The method of loci, also known as the mind palace, is a memory technique that allows people to recall information by associating the space they know well with the information they want to remember through the connection between object and subject. This memory technique primarily employs spatial memory to remember information efficiently. As another example of informational topologies, Yamada et al. (2017) developed a system called HoloMoL using the mind palace method to help users to memorize by combining information as mixed reality content with physical places.

The museum's potential is not solely derived from its exhibits, but rather from the interactions between the objects on display and the visitors. These interactions, along with the stories and communities that are formed within the museum, contribute to the vitality of both the collections and the visitors. The viewers' engagement with the objects endows them with meaning and with a sense of symbolic immortality. The museum also possesses a transformative power and an allure due to the mysteries contained within its unseen artifacts and untold stories. Malraux (1967) demonstrates how formal museums compromise the essence of works of art by clustering them together, diminishing the importance of individuals. He proposed a fictitious museum composed of photographs of works of art. Our society is evolving toward a more mediated culture.

Following the heterotopia concept, a paradigm shift in terms of contemporary senses has occurred within the museum structure. Online platform projects, such as Google Art and WikiArt, provide high-resolution images of selected artworks available to the public all over the world. Although there is a noteworthy difference between "Museum without Walls" and today's virtual museums on the Web, they have a similar conceptual purpose regarding transforming information and knowledge into forms that are available regardless of distance from it. The concept of the Digital Museum emerged in conjunction with the museum's expansion beyond physical places through the increased use of new media technologies. Various factors, including media and digital resources, influence the organization of virtual museums. According to Schweibenz (2019), the primary distinction between virtual museums and traditional physical museums is their level of accessibility.

With the adoption of digital media and a narrative focus, known as the "narrative turn," "the memory institutions" galleries, libraries, archives, and museums sector has undergone a transformation. Using technologies like virtual and augmented reality, 360-degree photography, and 3D reconstructions, they have transformed and created places and increased their use of digital spaces including websites (Basaraba, 2021). Immersion and interactivity in immersive environments aim to increase the sense of presence in a virtual environment; nevertheless, this does not mean that the digital environment is entirely composed of fictional elements. Reconstruction of physical artifacts into digital media uses 3D data acquisition methods, such as photogrammetry and laser scanning. These methods are frequently employed in virtual museums, virtual exploration, and cultural heritage contexts (Gonizzi Barsanti et al., 2015; Haydar et al., 2011; Pietroni et al., 2013). Hayashi et al. (2016) developed a virtual museum capable of displaying planar artifacts using web scraping to extract necessary information. The system allows users to choose the museum's content that is designed by authors. MR systems in museums can enhance the typical visitor experience by combining historical interactive visualizations with related physical artifacts and displays. MuseumEye application (Hammady, 2018) focuses on different guidance techniques to improve visitors' experiences. Providing adaptable exhibition spaces, Komianos et al. (2018) provided automated virtual exhibition construction based on adaptive exhibition topologies. They state that the facilitation of visitors' navigation can enhance users' visiting experiences.

Virtual museums and immersive cultural heritage studies can be interpreted as heterotopias of deviation and crisis. The risk of extinction due to natural or unnatural factors has created a need for the reconstruction of artifacts in a digital medium. The deviation has been occurring due to technological advances that have redefined almost every aspect of various practices. Additionally, COVID-19, as a worldwide crisis, created boundaries between people and places, which accelerated virtual museum studies to provide accessibility.

Place-making activities have been provided by games for decades, where players can shape and influence the virtual spaces within the game. These activities can range from building and constructing structures and landscapes to customizing and decorating the appearance of these spaces (Basabara, 2021). For individuals, they provide an opportunity to express creativity and personalize their virtual spaces and foster a sense of ownership and attachment to the game world, as players feel a sense of pride and accomplishment in creating and shaping their own spaces. Boldi et al. (2022) explore the technologies of crisis through video games. According to their study, the COVID-19 pandemic caused routines to be confined to people's homes, leading some to feel disconnected from their usual places of habitation and to turn to video games as a way to escape. Some people used games to explore virtual worlds that were different from their everyday lives, while others used games to recreate or substitute for meaningful places that were no longer available to them. Researchers suggest that games or online communities could be designed to allow people to explore and customize virtual spaces to strengthen their attachment to them and recreate lost places to help people find new opportunities and meanings in their everyday environments.

### 2.3. Immersive Technologies

Virtual, augmented, and mixed realities have been widely studied and have seen a range of applications in various fields. While virtual reality has a well-established definition, mixed reality has been described using a variety of terms, leading to ambiguous definitions. Mixed reality combines elements of both augmented and virtual reality, with the degree of combination determining where it falls on the reality-virtuality continuum (Milgram and Kishino, 1994). Virtual reality is a technology that uses real-world visual perception in artificial computer-generated environments through stereoscopic vision, providing an immersive experience with motion capture. Paul Milgram and Kishino propose the terms "augmented reality" and "augmented virtuality" to describe environments closer to the center of this spectrum. Augmented reality maps virtual elements onto physical space to create a hybrid of the real and virtual, often through smartphones. Mixed reality technology uses head-mounted displays to combine AR and VR capabilities, such as Microsoft's HoloLens.

### 2.3.1. Interface and Interaction

The user interface (UI) is the means of communication between a user and a computer system or application. Most personal computers use a graphical user interface (GUI) which uses visual elements such as icons, buttons, and menus and input and output devices like a mouse and keyboard to display a screen. Different types of interfaces include form filling, which involves the user entering data into a pre-defined form or template, and direct manipulation (Shneiderman, 1983), where the user interacts with on-screen objects in a way that directly affects their behavior. Command languages are sets of instructions or commands that a user can enter to interact with a computer or software program, often used in text-based interfaces such as in a terminal. Natural language processing (NLP) allows users to communicate with a computer or software program using natural language rather than specific commands, often used in voice assistants and chatbots (Gilbert, 2019).

Immersive environments may require more interactive interfaces with specialized input devices such as motion controllers or haptic feedback devices. These interfaces can include head tracking, hand movement tracking, voice control, eye tracking, body movement tracking, and virtual hand manipulation. To effectively manipulate objects in virtual reality environments, interaction techniques should allow for object selection, positioning, and rotation. The design of these techniques, often involving hand manipulation, is important for the overall user experience of a virtual reality environment (Mann et al., 2022).

### 2.3.2. Presence, Embodiment, and Experiencing Architecture

The human body serves as a system for acquiring, processing, and displaying information from the physical world through the senses. Virtual worlds are digital environments that users can interact with using technology that differ from the physical world. They require technology to experience and interact with objects, spaces, and people. Presence refers to the sensation of being physically present in a given space and time, often experienced in virtual reality environments (Schultze, 2010). Factors that can affect the level of presence include the realism of the sensory stimuli in the environment, the believability of the environment as a real space, the user's level of engagement with the environment, and the user's prior knowledge and expectations. Virtual worlds offer a chance to examine the role of the physical body in the communication and the effects of communicating without a physical body (Spence, 2020). Merleau-Ponty's philosophy (Lehtinen, 2022) emphasizes the concept of embodiment, or how the body shapes and is shaped by our perception and experience of the world. He argues that the body is not simply a means of engaging with the outside world but also plays a role in processes of perception, understanding, and expression. The body is therefore central to our subjectivity and understanding of the world.

In "Experiencing Architecture," Rasmussen (1964) argues that the sensory experience of a building or space is crucial to our understanding and interaction with the built environment. He believes that architecture should be evaluated based on how it is experienced by the user, rather than just its aesthetic or functional qualities. Rasmussen explores how various sensory experiences, such as sight, sound, touch, and even smell and taste, contribute to our perception of a building or space. He explains that the way we experience architecture is based on the perception of solids and cavities, figure-ground relations, rhythm, scale, proportion, sound, lighting, color, and texture. He divides the process of creating and interpreting architectural forms into two categories: solid-minded and cavity-minded. Solid-minded architecture involves creating forms by combining solids or shells of voids, while cavity-minded architecture involves carving out components from a large solid to create spaces. Rasmussen describes architecture as "the art of playing with solids and cavities."

According to Luck (2014), the design process often involves behaviors that may seem incongruous but serve a deeper purpose. These movements, which pertain to the design of a structure, symbolize something and provide a fleeting visual representation of architectural concepts and the anticipated sensation of movement within the architectural form. This process is referred to as "aesthetic becoming," and reflects the creative process behind the emergence of these behaviors. The relationship between a building's physical form, how it is experienced, and the reactions it elicits are subjective, intuitive, and complex.

### 2.3.3. Usability, Acceptance, Trust

Technology acceptance refers to the willingness of an individual or organization to use a particular technology. By understanding the factors that impact technology acceptance, designers and developers can create more effective and appealing technologies (Rheingold, 1991). Usability is a measure of how easily an individual can use a system or product to achieve their goals. It affects the performance, efficiency, and satisfaction of the user, as well as their overall experience. Factors that contribute to the usability of a system include the cognitive abilities and limitations of the user, and the design of the system's interface, organization, and structure (Salanitri et al., 2015).

The concept of trust in technology has been studied extensively in the field of human-computer interaction (HCI). Previous research has shown that trust in technology is influenced by factors such as usability, technology acceptance, and presence and that these factors also affect each other (Lippert and Swiercz, 2005). Trust in technology is a multi-dimensional concept, and researchers have identified several dimensions that contribute to trust between users and technology. These dimensions include predictability (the ability of technology to adhere to previously established performance standards), reliability (the perceived dependability of technology in certain situations), and utility (the perceived usefulness of technology) (Mcknight et al., 2011).

### *2.4. Procedural Content Generation*

The automatic creation of digital assets through algorithmic means and patterns with little to no user input is known as procedural content generation (PCG). In the gaming industry, algorithms are employed to create complex items like road networks, buildings, living things, as well as landscapes and plants. PCG is being studied in fields other than computer science, interdisciplinary approach in the scientific sciences, including biology, architecture, urban studies, and psychology, is triggering rising interest in other communities, and it has become increasingly prominent (Prusinkiewicz and Lindenmayer, 2012). As the importance of PCG for production increases, researchers are exploring new ways to produce high-quality assets, either with or without human input; therefore, new paradigms have also risen. Various techniques have been produced through machine learning (PCGML) and deep learning, such as neural networks, auto-encoders, and deep convolutional networks; Markov models, n-grams, and multi-dimensional Markov chains; clustering; and matrix factorization (Summerville et al., 2018).

Designers are increasingly utilizing autonomous tools to complete complex tasks faster and in novel ways. However, the deterministic nature of these methods can produce similar results repeatedly and may not provide the same level of creative user experience as hand-crafted and designed content. Most recently developed techniques aim to create fully autonomous approaches. Many researchers have proposed methods for the automatic generation of floor plans and buildings, but only a minority allows for customization of the end product of the algorithm. Products that are not open to customization have been criticized in the field of architecture (Porter and Hanna, 2006). In this study, we aim to offer greater control over the system on demand by manipulating the autonomy level of the algorithms and providing a "design space" for the user to customize the end product. Therefore, we sought methods that offer various autonomy levels and design space to the user.

Previously, large-scale procedural modeling of virtual worlds typically only resulted in empty structures devoid of internal divisions and interconnections. Different approaches have been proposed to solve this problem, including tile placement, room enlargement, inside-outward extension, and subdivision. Using tile placement strategies, the domain is partitioned, and the resulting grid is filled with tiles representing the rooms. Without relying on the limits of a building as a constraint, inside-out algorithms distribute rooms in accordance with the requirements for room connection. Growth-based algorithms disperse room seeds throughout a building, and the rooms gradually fill the interior space and grow to their full size (Camozzato, 2015). A constrained growth-based method for floor plan generation was presented by Lopes et al. (2010). In their method, the sizes and positions of the rooms are defined by the user. The rooms are then expanded until they become square in shape, and the empty space is then filled with further irregular shapes by moving the rooms. Given that the floor plan can be divided into subdivisions, it is possible to connect adjacent rooms, and subdivisions with doors and windows are also generated for each room. They indicated that minimal complexity and effective data structures are the main reasons for the success of their methods. Graph Approach to Design Generation (GADG), a method for automatically creating rectangular floor plans based on an existing graph extracted from floor plans, has been proposed by Wang et al. (2018). They developed a mechanism to manipulate the rooms based on two transformation principles, addition, and subtraction, by mapping the floor layouts to connectivity graphs. Users can specify the maximum width-to-height ratio for each created room using the algorithm.

Building boundaries are used by subdivision algorithms to divide the interior space into rooms. A technique called binary space partition (BSP) uses hyperplanes to recursively divide a space into convex sets. This subdivision becomes a representation of the scene as a tree data structure known as the BSP tree. Baron (2017) examined many methods for creating procedural material that may be applied to both 2D and 3D projects. They created pairings by using five-room and corridor creation algorithms (Random Room Placement, BSP Room Placement, Random Point Connect, Drunkard's Walk, and BSP Corridors). For indoor contexts, Yang et al. (2022) presented a reconstruction strategy for room layouts. To create the optimum in-door polygonal models, they integrate voxel-based room segmentation and space partitioning. The technique involves room semantic data to divide subspaces.

Cellular automata (CA) is another technique used for architectural generation. Cellular automata apply a set of rules to cells, and each cell executes those sets of rules with respect to its neighbors. Cells can acquire a finite number of cell states. Herr and Kvan (2007) show how a variety of cell shapes and sizes used in one CA model may support architectural form findings.

Additionally, they proposed a theoretical framework for the integration of CA into the design process. Araghi and Stouffs (2015) explore CA systems for high-density residential building forms. Based on the capability of CA to create complex rule definitions, they also address solutions for architectural problems such as density, accessibility, and natural light via rule definitions. Cruz et al. (2016) discuss variations of classic CA cell shapes to derive a variety of architecturally feasible forms capable of generating aggregate spatial units to match specific spatial configurations.

## 3. Design Rationale and Structure of the Framework

Based on the literature regarding heterotopias and discourse, we generated a set of rules to provide consistency in methodology and a comprehensive framework structure. The system should (1) have dynamic power relations and be open to changing its functions to create resistance; (2) provide juxtapositions of multimedia elements on an "operating table" that allow certain interactions; (3) provide a ground for "self-formation" and a "focal point of resistance" through individuals who construct "authorities of delimitation" together with the system; and (4) provide a relationship with other places by either creating an illusionary space or an orderly space composing "grids of specification"; (5) able to create "liquid architectures" according to needs of the users that can convey spatial qualities that are perceived via embodiment.

Throughout the methodology, we emphasized end-user development and meta-design principles to empower users, create a design space, and encourage informal use cases for novice and experienced users. Therefore, to simplify the mechanisms and the features that we integrated to transform users into curators of their digital environments, they were carefully selected following the previous studies on multimedia, architecture, and immersive technologies. In the subsequent sections, implementation details for VR, MR, and PC technologies are given in accordance with the flow and the layers provided.

### 3.1. Virtual Museum as an Operating Table

Different form the previous studies, the virtual museum concept in this study is not referred to as a static "emplacement" in digital space. On the contrary, it becomes a dynamic structure, an operating table where the user operates on elements by including, excluding, and transforming them. Designing a virtual museum can be viewed as an infinite process that begins with documentation and continues with imagination and experience in the virtual environment, forming a multilayered discursive space. When personal data is categorized and stored in a digital space, it can be seen as a digital archive. If it is displayed or played, it is considered an individual exhibit; if multiple files are displayed within a context, it is considered a digital exhibition. As it provides a categorization of documentation and displays archives in the form of exhibitions via the built environment, virtual museum space can transform individual archives into personal exhibitions creating "grids of specification."

Through the experience, we used the terminology museum spaces to increase usability and decrease the complexity of the framework. We defined two different modes for the application, namely curator and visitor modes. Curator mode defines the active creation of heterotopias, while visitor mode excludes the control layers. The framework provides a system for users to store and display digital content, including text, images, videos, 3D models, and audio, in a way that allows for easy access and manipulation. To be able to categorize items, we named the assets according to museum conventions, such as painting, sculpture, etc., and added labels to provide more data. We used asset style, artist, location, and time data to provide users with automation in reordering systems of heterotopias. On the other hand, the system is open to manipulation and interpretation. For example, the four variables that are used for sorting and grouping can be systematically employed by the user for different variables.

### 3.2. Dialogue Between the User and the System

The hardware provided, the general structure of the system, the interface design, interaction, and navigational capabilities all directly affect the experience. For example, hardware defines the physical comfort level, interaction modalities, and technical borders. The dialogue between the interface and the user affects multiple features of the heterotopia created by the user and shapes the user experience. Through the interface, the user grasps the level of autonomy. Interface, interaction, and navigation provide the narration for the discursive space and sense-making processes.

#### 3.2.1. Programming Language and Development Environment

To eliminate the results stemming from hardware differences in VR and MR, an Oculus Quest 2 HMD system was used to stream the audial and visual content. With the feature called Passthrough, using the sensors on the headset allows seeing a real-time view of the surroundings together with the virtual content.

The main reason that we used the broader term mixed reality instead of augmented reality in this study is the interaction and anchoring capabilities of the device, which combines AR and VR. The Oculus Insight tracking system is developed by Facebook Inc. and made available on the Oculus Quest 2. Oculus Insight tracking relies on three streams of sensor data. Continuous streams of data provide hand gesture recognition, physical context awareness, and the position of controllers and hands. The Oculus Quest 2 Controllers as interactivity devices and hand tracking were used to allow the users to navigate through the virtual environment (VE). The buttons on the controllers were visible in the virtual environment, including the hand position. The Oculus Quest 2 tracks the head movement to present the correct virtual world image to the eyes (LaValle et al., 2014), and it analyzes the user's head movement in real-time to control the view, which enables natural interactions, leading to high levels of presence and immersion (Desai et al., 2014). The HMD includes an adjustable head strap that sets the hands free for the controllers. The combined weight of the HMD (503 grams) and controllers (126 grams/per controller) is 755 grams, which facilitates comfortable use for extended periods.

All versions of the framework were developed using Unity. Unity game development platform was selected because of its prominent features among the other opponents in the industry, and it is one of the natural development platforms for Oculus Quest 2. The Unity game engine is compatible with various platforms including desktop, mobile, console, web, and more. Unity supports C# programming language which is designed for Common Language Infrastructure by Microsoft. It is a general-purpose, multi-paradigm, and object-oriented language that allows developers to build applications that run on the .NET Framework. A programming paradigm called "object-oriented programming" (OOP) is based on the idea that "objects" can hold both data and code that manipulates that data. An object in OOP is a self-contained unit composed of both information and programming code. The code that operates on objects is created to reflect the behavior of real-world objects, and objects are used to represent real-world concepts or objects. OOP is characterized by the use of encapsulation or the bundling of data and the code that manipulates it into a single unit or object. Using encapsulation, the framework becomes open for transformation and additional features. This changeable structure gives the framework adaptability of the heterotopias. Besides Unity, the Microsoft-driven project Mixed Reality Toolkit (MRTK) provides a set of components and features used for the VR and MR parts of the application. The 3D models were created in RhinoCeros and Blender and enhanced by 3D modeling distribution web environments which provide copy-left models such as CG trader, Sketchfab, and TurboSquid.

### 3.2.2. Interface and Interaction

The interface is designed to guide the user through the experience and construction of heterotopias layer by layer. Different types of layers are provided via different planes and sequentially presented. For the interface design, we preferred techniques especially suitable for novice users. For example, although the command line is very flexible since the learnability of command languages is generally very poor and not suitable for non-expert users, we did not include any command language interaction. The form fill-in interaction style was aimed at a different set of users than the command language, namely non-expert users. We used a form fill-in type of design for the tasks that require several data entries from the user. For example, the size properties of generation algorithms (Figure 1) and URL for connection are given in the style of form. To guide the user via the predefined rules and simplify the data entry, instead of using empty input fields, we preferred to use sliders to define and show the range directly to the user when it is applicable.

Menus are collections of options that are displayed on the screen, and, upon selection and execution, affect the system's state. The user chooses a command from a predetermined list of commands listed in menus using a system based on menu selection. Although it can slow down the frequent users, we prefer to use menu interactions in most cases since it is ideal for novice or intermittent users. To increase memorization and eliminate language boundaries instead of labels, we used icons for the command/menu items. To decrease the information overload and visual clutter, the items were clustered into sub-menus according to their functions (Figure 1). In the VR and MR versions, if the feature is not toggled off, menus follow the users as they move. In case of losing the main menu, a hand menu comprising the restart, quit, and call main menu options was included, which can be called via a hand gesture and a button on the controller (Figure 2). The 3D menus designed for VR and MR also allow users to interact via touch.

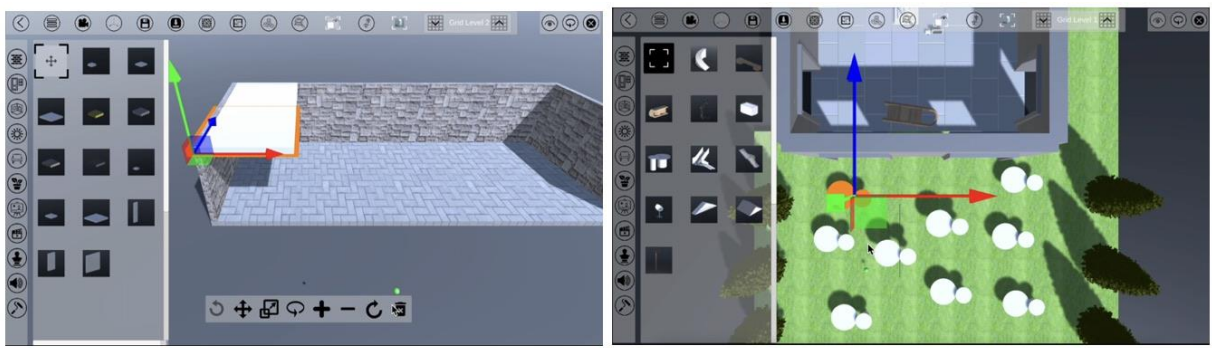

Figure 1. Example Interface options of the PC version.

We employed the direct manipulation principle for all objects that are placeable to the grid and 3D interface elements. Selected objects are highlighted and can be directly manipulated via hands, controllers, or mouse and keyboard according to versions. Direct manipulation is especially important for the embodiment aspects since it requires movement of the body. With this division, we aim to prevent unintended changes in models and remove unnecessary UI elements from the vision of the user, which, otherwise can disturb the users' experiences. The user experience for all versions starts with a welcome screen and continues with a panel that is used for importing the digital archive. The archive that is analyzed and environmental elements to construct places are provided to the user with a series of panels to construct the discursive habitat. While the PC version uses the keyboard and mouse as input, for VR and MR versions, we offer controllers and hand interaction. The gestures are components of human-computer interaction that have become the subject of attention in multimodal systems. In this study, we used six hand gestures, namely point, select, release, teleport, call hand menu, and call main menu (Oculus). Hand interaction is provided with far and near interaction options. For all versions, the user can manipulate the elements including 3D UI elements, via scale, rotate or translate operations.

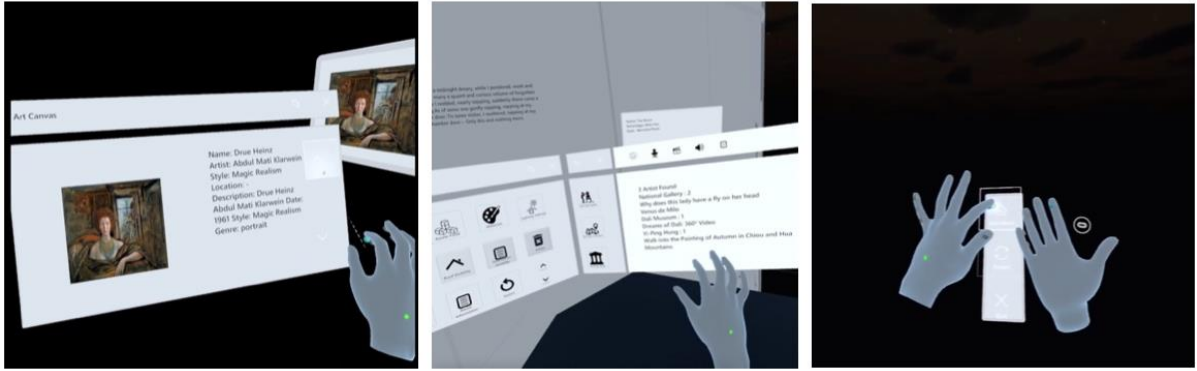

Figure 2. Hand Menu and Interfaces in VR environment.

### 3.2.3. Navigation

In authoring personal museum experiences, the users should be able to fully control their navigation in the museum and they should be able to freely explore and manipulate the objects to create their own virtual experiences in an interactive and flexible way. In the curation process, the user is not just an observer but can interact with the exhibits via constructive dialogue. The overall process and navigation of the user in the virtual museum enhances the understanding and keeps the interest alive by enriching the aesthetic sensitivities.

During the design process, we tried to make sure that all versions were as similar to each other as possible. However, inevitably, we used different methods in terms of navigation and interaction between the PC version and other versions. The PC version includes four types of cameras: first-person, top-down, isometric, and isometric bird view. All cameras can be controlled by W, A, S, and D buttons, and in the top-down view, zoom-in/out options are available with a mouse scroll. Also, there is a transport toggle in the control menu which allows the user to teleport all cameras in the scene with the left mouse click. The gestures are components of human-computer interaction that have become the subject of attention in multimodal systems. Gestures are captured through cameras on Oculus Quest 2 and can recognize various gesture types. Inside the VE, users can navigate, walk or teleport themselves using controllers or hand gestures. The tags are used to define colliders that allow teleportation. Objects with floor and ground tags allow teleportation. To eliminate motion sickness, we did not add movement via controllers.

### 3.3. Content Layer as an act of Self-Formation and Archival Practices

Being a force of resistance, individuals should determine the elements of the "surfaces of emergence" as an act of self-formation. Archival practices as a method of analyzing the discourse and its elements comprise the actions of self-formation on an operating table. To provide automation for self-formation activities, we constructed a pipeline that includes importing individual archives and transforming them into operable artifacts. We also provided an analysis of the archives based on grids of specification of museums.

### 3.3.1. Importing Files and Authority of Delimitation of the Framework

The first layer of delimitation of the elements will be represented on the operation table defined by the framework in terms of media types. As the elements of multimedia language, five types of assets and their museum equivalents have been identified. Texts in .txt, images in .jpeg and .png format, 3D models in .obj and .fbx format, audios in .mp3 format, and videos in .mp4 format are acceptable for generating the artifact displays. Within these constraints, being the focal point of resistance, the users are enabled to create a personal collection of 3D digital exhibits according to their interests and preferences. For the PC version, a custom file browser is written, which allows the user to choose the folder consisting of the files that will be uploaded to the application. For the VR and MR versions, instead of using a file browser due to the security properties of the device, the application pulls the data from Google Drive. To provide this feature, we include the virtual keyboard provided by Oculus and an input field where the user can write only the file ID of the data file. Using the ID, a direct link for Google Drive is constructed for a data file in .JSON format. The data file needs to include the file IDs of the objects to be downloaded into the database of the framework.

### 3.3.2. Artifact Classes

In addition to assets, the user can provide a data file in .CSS or .JSON formats which include certain features of the artifacts. Data columns are defined as the name, artist, style, location, time, size, and description. Five different classes were constructed inheriting from the base class according to acceptable asset types, which are paintings, sculptures, videos, sounds, and text. Each art object has common properties, such as name, description, artist, and style. According to the class, different properties are attained. For example, paintings and sculptures have location and size properties. According to file extensions, the tool copies the files into the data folder and groups them into artifact groups. If the data provided by the user is inside the folder, data is parsed into columns and for each line, a new art object instance is constructed according to its class. Algorithm searches for matching names comparing the data and files and constructs dictionaries and separate lists for the art objects. If no data file is provided, instantiated objects use the filename as a name and are placed without a description panel and with a generic scale. After importing processes according to extensions of the files, buttons are constructed for each file under the corresponding panel to spawn the artifacts in virtual environments. For the paintings, sprites of the buttons use the file directly, for other types of icons indicating the type are used. The name of the button is generated according to the name of the files. The placement of all objects will be explained in the grid system.

### 3.3.3. Artifact Holders

Following the end-user development, the system should provide automation to transform the assets of the individuals into artifacts that are open to manipulation so that users can benefit by making changes, plans, and experiments with the artifacts and finding the best position for them both in the virtual and the real environment. Based on features of OOP, for each artifact type, an artifact holder is designed which contains several scripts and elements that are necessary for the interaction. Holders are higher entities in the hierarchy, placed under the manipulation of the object's position, rotation, and scale. Each holder consists of a label panel where the short descriptions are placed, the Game object that will contain the assets, the Details button to open the ArtCanvas panel, a collider to manipulate, and a grid object collection script to fix the space between artifact and label. If the dimensions of a painting or sculpture asset are provided, the container object of the asset is scaled according to data, and the space between the artifact display and label remains stable. Additionally, each artifact holder has asset-specific features. For example, holders of video and sound have a play and pause button (Figure 3).

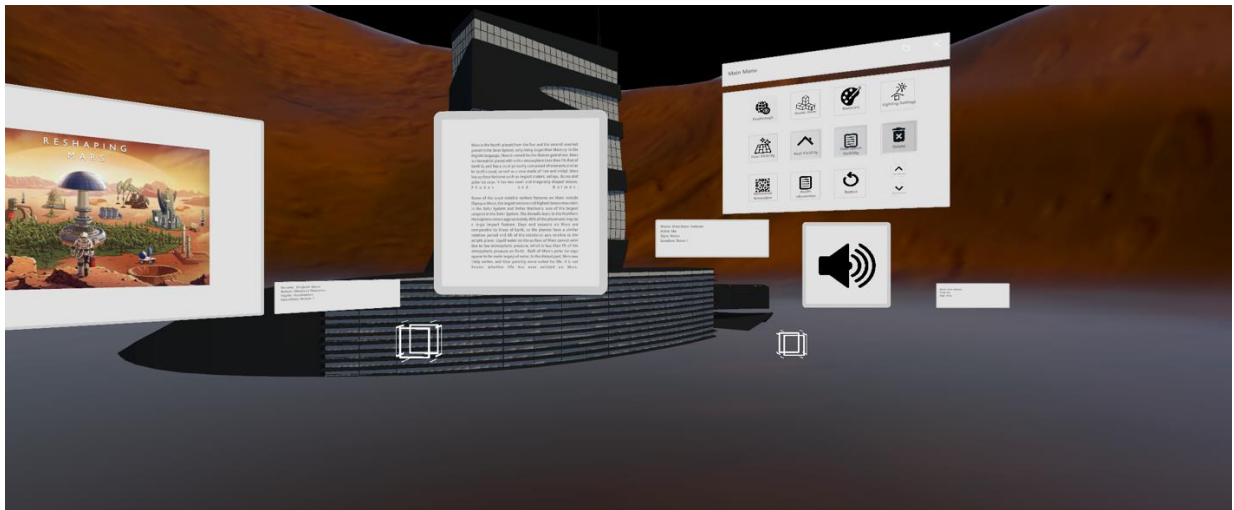

Figure 3. Artifact Holders in VR Environment.

For each class, artifact holders are designed as prefabs. When an artifact button is pressed, an artifact holder is constructed according to the related prefab. If data is provided, the name, artist, style, time, and location columns are given on the label. Also, we provide an ArtCanvas panel where the information in the description column is given with the artifact. The video, text, and painting artifacts are visually represented on the left side of the panel. For the sound and sculptures, icons are placed. On the right side of the canvas, the details section is provided. Users are also able to see a list of all assets they uploaded as a list in the artifact analysis panel where the information is provided based on the categorization of artifacts.

### 3.3.4. Analysis of Artifacts and Grids of Specification of the Framework

The first classification of the assets is performed according to formats to provide an accessible archive. If the data is provided, it is parsed into columns and for each class, artifacts are grouped according to four specifications. The framework offers three strings and one integer for archival indexes which are named in accordance with the terminology of museums and named as artist, style, location, and time. The constructed lists are provided through an artifact analysis panel. These groups are also prepared for the PCG algorithms to calculate the dimensions of the rooms according to grouping options. On the other hand, based on the data types, users can generate different types of specifications.

### *3.4. Spatial Layers as an Act of Self-Formation and Architectural Elements*

The primary purpose of using architectural elements in a virtual museum context is to reorder the multimedia elements constructing spatial grids of specification. In this study, the idea of a "virtual museum" does not necessarily indicate a traditional built environment. On the contrary, it extends the space and forms a "liquid architecture" around the user. Every digital object that is placed, architectural or not, becomes information. Beyond being a partition for artifacts we offered place-production practices to provide "aesthetic becoming" experiences as an act of self-formation.

The layers of space in a heterotopia demonstrate how complex and multifaceted these spaces are, with different meanings and functions. These layers can be thought of as different levels of reality that are provided by various technologies. One layer of spatial meaning in a heterotopia is its actual physical location in the world and its meaning within context. PC, VR, and MR technologies provide different sets of relations with places and place production. The relationship with the physical location is provided via mixed reality technologies. According to the context, the user can transform the physical space with digital elements or create different places that are completely digital. Heterotopias have layers of spatial meaning that relate to the way they are experienced by individuals. Through embodiment and telepresence, technology provides users with endless possibilities. In this study, we provided architectural practices to increase the transformative effects of heterotopias, and through embodiment and presence, we aimed to present an operating table for those practices. To be able to provide an easy-to-use generic framework, we examined the architectural studies to identify essential features. According to Rasmussen's research in "Experiencing Architecture," various factors such as the perception of solid and cavity forms, the distinction between foreground and background elements, the size and dimensions of architectural features, the appropriate use of proportion and scale, the acoustics of the space, the lighting, the textures, and the rhythm all contribute to the overall sensory experience of a building or space. Therefore, we aimed to provide those elements in a technological landscape.

### *3.4.1. Construction Elements*

For the construction of architectural aspects, we identify the main construction elements as walls, floors, roofs, windows, stairs, and doors. Additionally, we add two different classes: landscape elements and furniture. All classes are provided in different panels. The main elements are available as default objects, and they are editable in terms of position, scale, and rotation.

### *3.4.2. Textures, Materials, and Colors*

Inside the Unity engine, 2D textures are tiled to provide materials. For the wall, floor, roof, door, and window objects, different textures are provided. For the consistency of place and ease of use, generic objects in the same layer share the same material properties. We added sliders that control the RGB values of materials that are used by the construction objects. Additionally, we provided different skybox materials for VR and PC versions.

### 3.4.3. Scale

Although all objects are scalable, we add different scale options to increase usability. The same wall, floor, and roof tiles are also provided in different scales, such as 1x1, 2x2, and 4x4. At the beginning of the experience, the framework provides two different scale modes for VR and MR environments; human scale (1:1) and model scale (1:20) (Figure 4). All models are automatically scaled according to the selection. Additionally, vertical construction components are scaled according to the selected grid height.

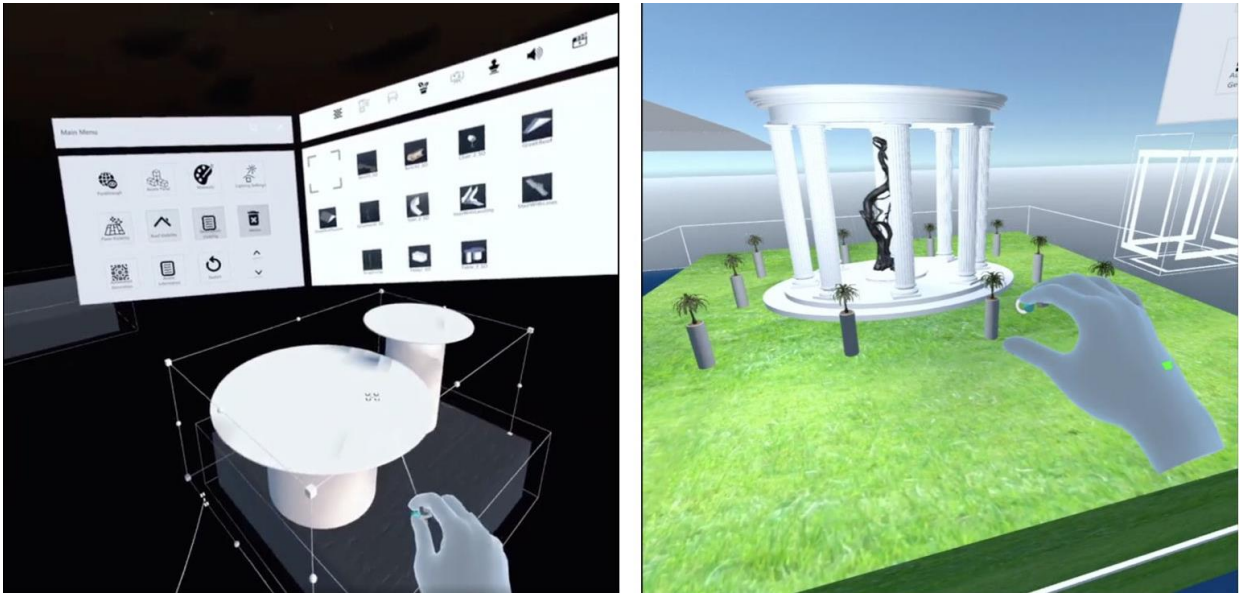

Figure 4. Human-Scale and Model-Scale Interaction.

### 3.4.4. Lighting Settings

Three different light sources are used in the project. Sunlight, as directional light, ceiling lights as light strips, and art objects have their lighting as a spotlight. In the lighting settings panel, each lighting type has a toggle that can be turned on and off (Figure 5). A slider is added to control the temperature of the light with values in Kelvin which are converted to the color of light.

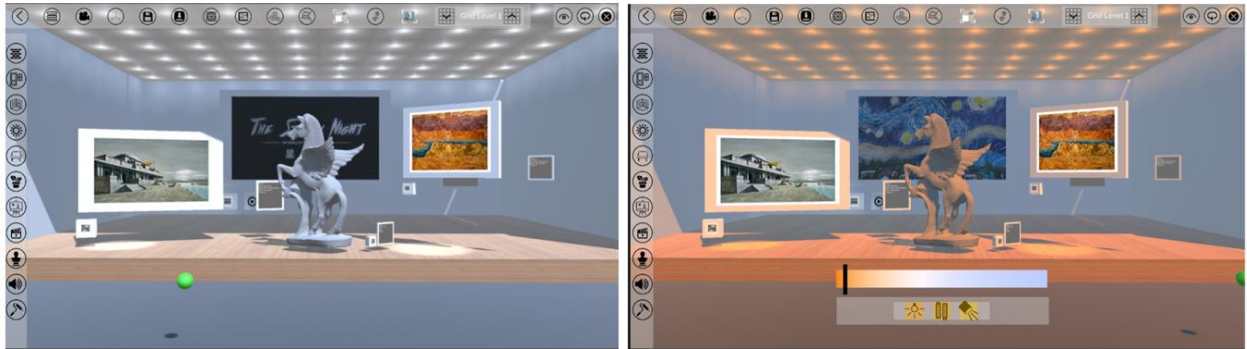

Figure 5. Artifacts under neutral and warm temperatures.

### 3.4.5. Placement and Grid System

The grid system was constructed to provide placement and snapping to correct positions to avoid overlapping. Users can set grid level and grid height properties. Grid level defines the number of floor levels, and grid height defines the height of the walls and the distance between two grid levels. For all placeable objects, including artifacts, three different Scriptable Objects (SO) were created. The SOs are types of assets in the Unity engine that allow saving data in an asset file rather than as a mono behavior attached to a game object. This allows data to be stored in a way that is independent of any specific game object and allows data to be easily reused and shared between different parts of the framework. The first type is floor objects which can only be placed on the layers identified as ground. The grid system is used for snapping floor objects to correct positions and holding the position data of floor objects. The second type includes vertical elements that are placed edges of the floor tiles such as windows, walls, and doors. Each floor object includes four colliders that are placed on the corners. This way, during the placement of the vertical objects on floors, the grid system snaps and rotates the vertical elements according to the edges of the floor objects on collision.

The third type includes other elements and artifacts which are placeable in every location. All SO types include two variations of one prefab, which are ghost objects and placed objects. The initial code on the scriptable objects adds the additional scripts that are necessary for interaction and hierarchy according to types. When the instantiation button is pressed, first, a ghost object is spawned which follows the pointer in MR and VR, the mouse position in the PC version. To place the object, the left click of a mouse, the select gesture with the right hand or right trigger is used according to the version. To cancel the selection, the right mouse click, and the select gesture with the left hand or left trigger button is used. For the delete operation, a toggle is added which allows users to delete the selected objects when it is on. If the model scale is selected, the grid system is placed on a table and every spawned object becomes a child object of the table. This provides a conversion of the scale and movement of the objects with the table.

### 3.5. Architecture and Archives Combined: Automated Generation of the Built Environment

The algorithms provide reordering mechanisms in 3D dimensional spaces constructing grids of specification based on archival data while offering fast prototyping to architectural practices. Based on discussions in architecture, heterotopias, and meta-design principles regarding power, autonomy, and control, we adapted the algorithms which can offer design space and control to the users (Figure 6). According to the requirements of this study, algorithms are able to generate content based on user data, and generated content is open for customization.

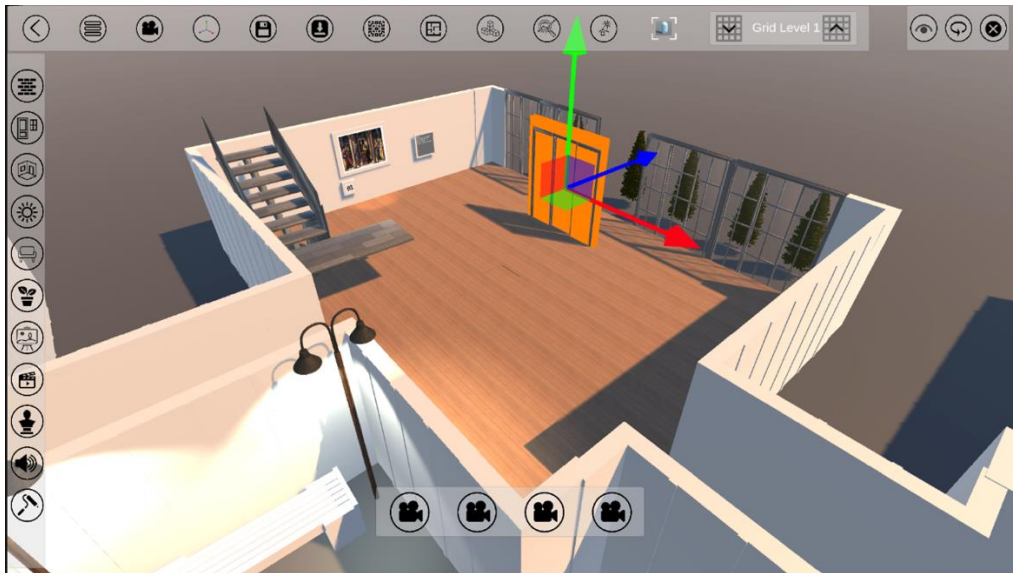

Figure 6. Transformation of Binary Space Partitioning Algorithm's output.

Rasmussen divides the production and interpretation processes of architectural forms into two: solid-minded and cavity-minded. One can start with a simple framework and add to it, or one can carve out certain components from a large solid to start. Drawing from this separation, we defined two scale options namely, room-scale and building scale, and employed one algorithm that can produce content based on dimension, window, and door requirements for room generation where the user can identify the dimensions of the solid and the number of cavities. We have provided two different generation techniques for building scale production. We created a solid-minded generation combining Binary Space Partitioning (BSP) and Cellular Automata (CA) algorithms and a constrained growth algorithm to provide cavity-minded production. While these algorithms are mostly used with random number generators to create different game-level designs, we injected user preferences and user data to create levels of automation. Each algorithm provides three levels of automation. Users can prefer to generate a 3D environment based on data provided, with random numbers, or they can define the parameters through the interface. This approach offers creative control over mechanisms that presents different levels of granularity. In all three cases, we made architectural choices that would ensure coherence in the environment produced to create a consistent user experience. Users can re-design the produced models of a building by using the same operation that they can perform on other assets. Without the scale operation of the user, the height of all vertical components is defined by the parameter given by the user at the beginning of the experience. To this end, as an input to algorithms, we used the same assets that we provided as main construction objects.

### 3.5.1. Calculation Strategies Based on Data

To be able to calculate the number of rooms and minimum dimensions, the data provided by the user is parsed into columns (Algorithm 1). According to style, artist, and location data, artifacts are grouped and provided via an artifact analysis panel together with artifact types. The number of rooms is defined according to the categorization selection of the user. For example, if the style is selected, the number of styles also defines the number of rooms. In this way, each artifact is assigned to a certain room. Since the dimensions of audio, video, and text artifact holders are already defined, their dimensions in the X-axis are directly added to the sum. Dimensions of the paintings and sculptures are extracted from data and if they are not available, the default dimension in the X-axis is defined as one meter.

| | **Algorithm 1:** Room Number and Size Calculation |
|---|---|
| 1 | Parse user data into columns |
| 2 | Group artifacts by style, artist, and location |
| 3 | Define the number of rooms based on user categorization selection |
| 4 | Assign each artifact to a room |
| 5 | **foreach** Room **do** |
| 6 |   **foreach** Artifact in a room **do** |
| 7 |     **if** the artifact is audio, video, or text **then** |
| 8 |       Add dimension in X-axis to sum |
| 9 |     **else** |
| 10 |     **if** Dimensions are available **then** |
| 11 |       Extract the X-axis dimension from the data and add it to the sum |
| 12 |     **else** |
| 13 |       Set the default dimension as 1 meter and add it to the sum |
| 14 |     **end if** |
| 15 |     **end if** |
| 16 |     Add constant label and space sizes to sum |
| 17 |     Divide the X-axis dimension by four and add to the sum as spacing |
| 18 |   **end for** |
| 19 |   Add minimum entrance size for two parallel walls (2-meters x 2) to sum |
| 20 |   Divide the sum by 2 to get the sum of two perpendicular walls |
| 21 | **end for** |

Additionally, the sizes of the labels and the space between displayed objects are static which is added to the total sum for each artifact. To create lists, artifacts are queued according to the date aspect. The artifacts that are not provided with a date are added according to alphabetical order. From the constructed list, space between each artifact is calculated based on their sizes and added to the sum together with the minimum entrance size for two parallel walls. In the final step, the calculated length is divided into two which gives the sum of two walls that are perpendicular to each other. Calculated lengths are provided to other algorithms as a parameter that will define the dimensions of the rooms.

*3.5.2. Constrained Growth Algorithm*

A procedural method's capacity to define its limits depends on space, allowing it to precisely use the region it generated as input before continuing to generate within it (Algorithm 2). Using a building outline for a floor plan generation allows for determining the borders of the procedural method. To provide consistent footprints for users, we searched for floor plans of existing museums. Using the websites such as Archdaily and Divisare, we obtained 20 different floor plans that have different complexity levels in terms of space. Using OpenCV, we extracted the outlines of the floor plans with filling operation, and we transformed them into footprints. We used footprints as the solids that will hold the cavities that will grow inside them. After several experiments, we defined the sizes of .png files that contain footprints as 128x128 pixels due to computation limitations. The prepared footprints according to pixel and color requirements are provided as options to users. To operate the algorithm, the user first selects the footprint to work on.

Depending on the preference, the user can define the starting growth points by selecting the pixel form image of the footprint or if data is available, the user can choose the option from sorting categories (style, artist, location) that defines the number of rooms, and the algorithm selects the random points according to optimal distances. If there is no data available, the user can also prefer the algorithm to select the number and location of the points randomly. The starting points form a square with a 1-pixel void and 8 pixels around as boundaries (Figure 7). After the selection of starting points based on pixel-based search, starting points begin to grow until they reach the borders of the footprint or borders of each other forming cavities inside the footprint. When the growing process is completed, corners of the containers are found by continuously adding and comparing the positions of the pixels which are owned by a certain room. According to corners, center positions, and length of the walls are defined to place 3D wall tiles. For each room, the 3D floor and their mirrored version of roof tiles are placed according to empty pixels and walls.

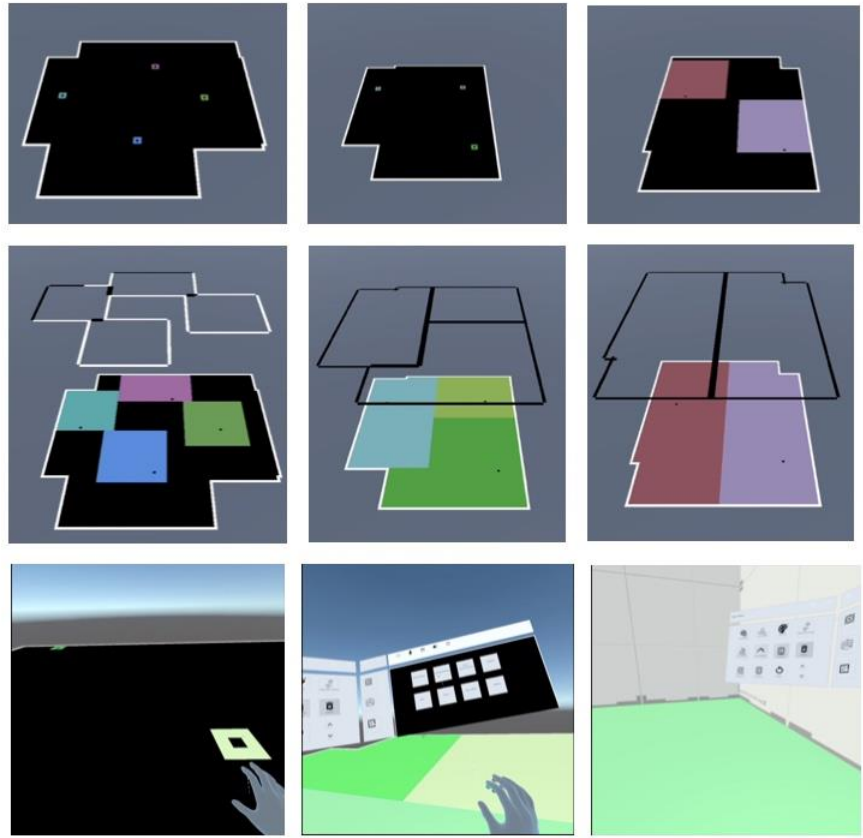

Figure 7. Example Outputs of Constrained Growth Algorithm.

### 3.5.3. Room Generation

In the room generation algorithm, a cubic room is generated using the generic wall, roof, and floor assets as a solid. For the data-based version of the algorithm, the user can select the sorting options for the assets and each room according to the number and sizes of the artifacts that the room will contain, the width and depth dimensions of the room are calculated, and the initial room is scaled according to calculations. For each room, a button is produced to spawn the room in the virtual environment. In the version based on user preferences, initially, the user places a generic room that holds the generation algorithm. The user can play with the cavities and solids by defining the dimensions of the room, the number of windows, and the number of doors. The assets that are used to construct the initial solid together and window and door assets are given to the algorithm. When the user presses generate button; first, floor and wall tiles are replaced according to the given dimensions. Then, the algorithm changes wall tiles to window and door tiles according to the given numbers by the user (Figure 8).

| | **Algorithm 2**: Constrained Growth Algorithm |
|---|---|
| 1 | Initialize new list cornerList |
| 2 | From (minX, minY) search four directions for a boundary pixel |
| 3 | **if** it is found **then** |
| 4 |     Add to the cornerList |
| 5 |     **else** |
| 6 |     Move |
| 7 | **end if** |
| 8 | Sort corners clockwise |
| 9 | **if** a room is placed, **then** |
| 10 |     Check Region |
| 11 |     **if** the region is clean, **then** |
| 12 |         Draw room boundaries |
| 13 |         Add a room to the room list |
| 14 |         **else** |
| 15 |         Terminate the Room |
| 16 |     **end if** |
| 17 | **end if** |
| 18 | **if** it is not paused, **then** |
| 19 |     Iterate through the list of rooms and find candidate walls |
| 20 |     **while** the number of growable walls > 0 **do** |
| 21 |         Grow room |
| 22 |         Update walls |
| 23 |         Update texture |
| 24 |     **end while** |
| 25 | **end if** |

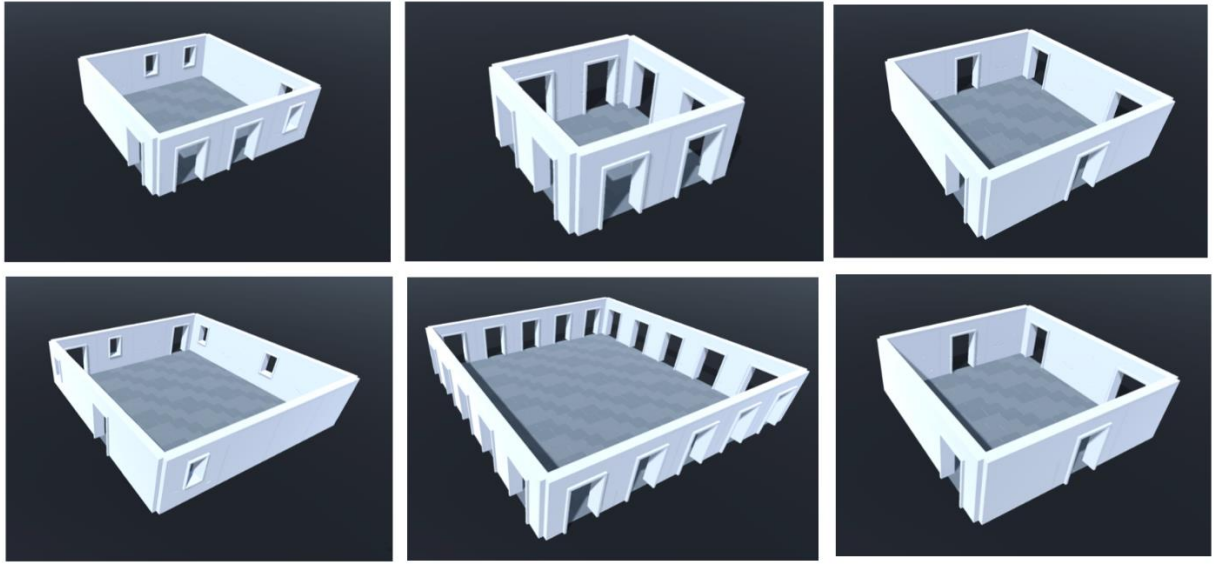

Figure 8. Example Outputs of Room Generation Algorithm.

### 3.5.4. Binary Space Partitioning and Cellular Automata

Cellular automata (CA) and binary space partitioning (BSP) are two different algorithms that can be used separately or in combination to generate layouts. BSP is a technique used in computer graphics to divide a 2D or 3D space into smaller subspaces, or nodes. In general, the number of subspaces of BSP is defined with a random number from a range with a seed provided for variation. It is mostly used for generating complex 2D dungeon patterns. This study used BSP to produce interconnected rooms and corridors; however, different from the most common techniques, instead of using random numbers we constrain the algorithm via user preferences.

A type of mathematical system known as a cellular automaton is made up of a grid of cells, each of which has a limited number of possible states. According to a set of rules, each cell's state is determined by the states of its neighbors. This means that by interacting with one another in cellular automata, simple rules can be used to create complex patterns and structures. In video games, CA is typically used to create more natural patterns like caves and forests. In this study, we created a 3D model of the generated grid data based on the cell state using CA. The basic idea of BSP is that any plane can divide space into two spaces. Therefore, it creates different areas by recursively dividing the level space. Any half-space that we continue to define a plane in will be further split into two smaller subspaces. Subspaces are created and a spatial binary tree is formed through continued division. The divisions may be placed at random points so that not all areas are of the same size and shape. Once the areas are a suitable size, each area is converted to a room, and connections are added between adjacent regions. In the data-based version of the BSP, the number of rooms and their dimensions are provided by a data analysis algorithm that calculates parameters according to the sorting option selected by the user. Other variables are randomly produced by the algorithm and altered via seed. In the absence of data, with the defined seed, the algorithm randomizes all parameters used for the generation. For the version based on user preferences, the user can define maximum and minimum values for dimensions of corridors and rooms, dimensions of footprint, and the number of rooms so that the user can give the parameters of the solids.

We generated a layout using BSP to start by defining the initial boundary which is the footprint that has dimensions depending on the user's preferences (Algorithm 3). Given sizes for the footprint can change the overall placement of the layout. For example, if the user prefers to create longitudinal side-by-side blocks or more compact structures, it can be done by playing with the width and depth ratio. The next step is to divide this footprint into smaller spaces in accordance with the number of rooms and dimensions defined by the user (Figure 9). In binary space partitioning, a seed is a starting value that is used to generate a random sequence of values. The random sequence is then used to determine the layout. We used the seed to randomly select the location and orientation of the division to create variety. The seed value is used as the input to a random number generator, which produces a sequence of random values based on the seed. By using the same seed value, it is possible to generate the same layout multiple times. The divisions are created based on the room and corridor dimensions and seed value which continues until the number of rooms defined by the user is reached.

Each node in the constructed tree structure represents a subspace within the layout. To create the actual layout of the tree, the tree is then traversed, and at each node, if the node can create a room, it is added to the layout and the algorithm tries to place a corridor in all four cardinal directions. If there is enough space, the algorithm places a corridor and moves on to the next node. If there is not enough space, the algorithm terminates the process and restarts to division. The BSP algorithm operates on cells, which makes it suitable to combine with the CA for a rule-based 3D generation. We defined several states for the cells to construct the 3D model of the generated data structure. Before the division, all the cells are in the state of empty when all the rooms take place in the spatial binary tree cells change their states to the wall, corner wall floor, door, and window. While wall and roof tiles do not need any rotation, for wall, corner wall, and door tiles there are different states which apply the rotational operations. Applying the rules, 3D assets take their places. For the roof, instead of creating a new state, we injected the responsibility to the cells in the floor state, which can create roof tiles by duplication according to the defined height.

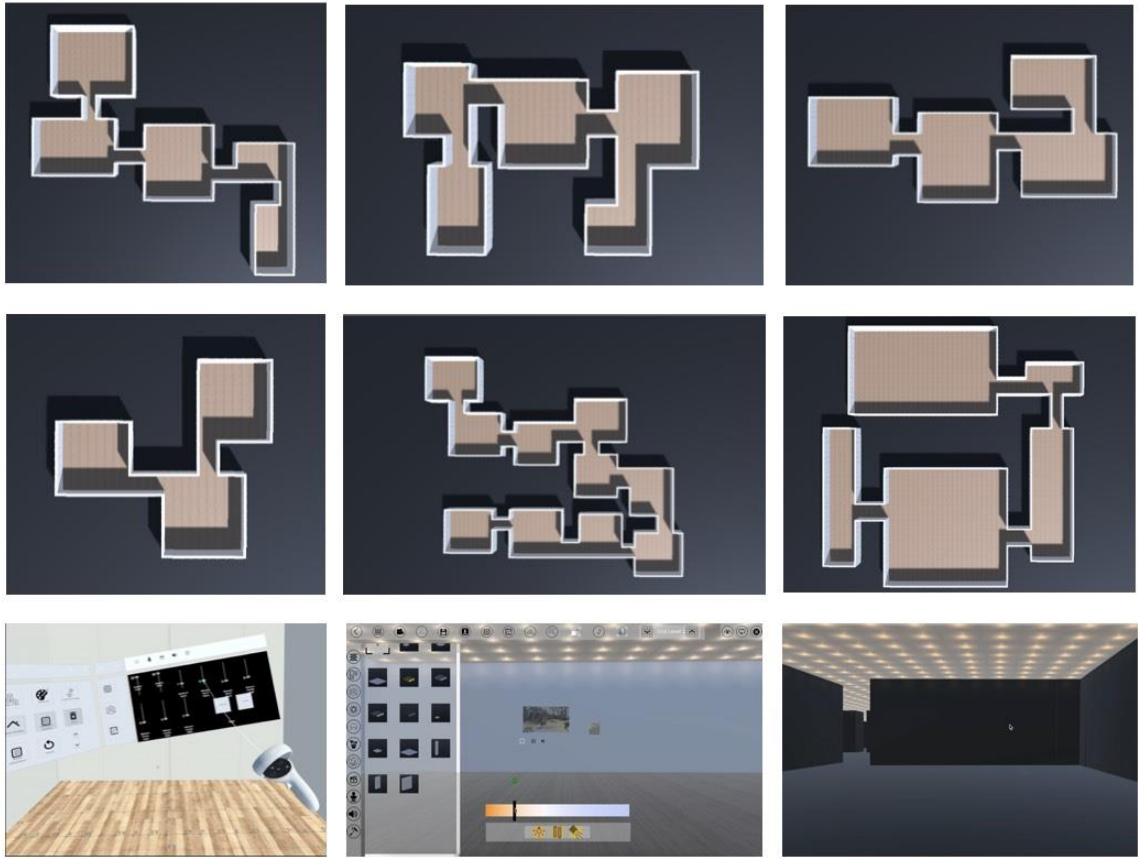

Figure 9. Example Outputs of Binary Space Partitioning and Cellular Automata.

## 4. Evaluation

### 4.1. Ethics

Ethical Approval of Research was approved by the Middle East Technical University Human Subjects Ethics Committee with the approval protocol code of 0244-ODTUİAEAK-2022 in April 2022. Before the start of the study, every participant signed an informed consent form. The form includes a brief description of the study's aims and objectives, the importance of participants' input, data collection methods, possession of participants' personal information, the intended use of their data, and what is expected of them. Participants had been warned about the potential negative effects of VR platforms, such as motion sickness and nausea, and it was explained that participants had the right to leave at any time if they feel uncomfortable or disturbed during the experiences. All the information collected for this study is anonymized and figures that may include personal data were excluded. No compensation was provided to participants.

| | **Algorithm 3:** Binary Space Partitioning and Cellular Automata |
|---|---|
| 1 | Initialize the layout with the given footprint dimensions |
| 2 | Use the seed value as the input to a random number generator |
| 3 | **while** layout.num_nodes < user.num.rooms **do** |
| 4 | Select a random location and orientation for the division |
| 5 | Divide the layout into smaller subspaces |
| 6 | **end while** |
| 7 | Traverse the tree structure of the layout |
| 8 | Recursively visit each node in the tree structure of the layout |
| 9 | **if** the node can create a room, **then** |
| 10 | Add it to the layout |
| 11 | Try to place corridors in all four cardinal directions |
| 12 | **else** return |
| 13 | **end if** |
| 14 | **if** layout.num_nodes = user.num.rooms, **then** |
| 15 | Iterate over the cells in the layout and place 3D models according to the states |
| 16 | **else return** |
| 17 | **end if** |

## 4.2. Participants

A total of 30 participants were recruited. Recruited respondents were between the ages of 18–35, with a mean age of $22.52 \pm 4.13$ years. Each participant had a normal or corrected-to-normal vision (self-reported). The scope and focus of the research topic, the methods utilized to collect the data, and the amount of information that was acquired from each participant are all factors that can affect the sample size (Braun and Clarke, 2019). Enough information should be gathered to adequately address the research questions. Given the numerous methods of data collection and the depth of information gathered from the participants, 30 participants were an adequate number, and the analysis from which produced important insights.

## 4.3. Procedure

We conducted our evaluation using a quantitative survey in the form of questionnaires. Additionally, semi-structured interviews were undertaken as an exploratory method to provide more detailed and in-depth insights about the use of the application and preferences. Before the version test, participants were asked to fill out the Gamer Motivation Profile (Yee, 2016), Immersive Tendencies (ITQ), and Tool Competence questionnaires. The users were provided with identical data and museum collections to avoid inconsistencies resulting from the data and media. Following the procedure, participants were given access to their collections. Users were first asked to freely use the application, then test the specific features, such as lighting settings and procedural generation algorithms. Once participants had used all three versions, the participants were given the questionnaires intended to assess several aspects of the versions. For the qualitative evaluation of the framework, a widely used System Usability Scale (SUS) (Brooke, 1996), Presence Questionnaire (PQ) (Witmer and Singer, 1998), and Technology Acceptance Model (TAM) (Venkatesh & Davis, 2000) questionnaires were adopted and distributed to the participants. In addition to those, we composed three questionnaires for lighting, procedural generation algorithm, and interaction preferences. Later, a short semi-structured interview was conducted. Each session lasted about one hour.

## 4.4. Data Collection Methods

### 4.4.1. Semi-structured Interview

After all the platforms had been used by the participants, the semi-structured interview was held after the session to gather further information about general impressions of using the versions. The guide consisted of 11 questions that focused on participants' interpretation and perception of the framework, experiences of using the different versions how they could be improved, and which versions and options they preferred and why. Additionally, informal conversations were made to identify possible factors that had been overseen, such as mistakes in applying the method and software errors. In this interview, they were also asked to describe their experience using the application in their own words.

### 4.4.2. Questionnaires

The System Usability Scale is a 10-statement self-report scale that is used to analyze and study a system's usability for general evaluations. It is assessed on a Likert scale of 1 to 5, where 1 is for strongly disagreeing and 5 is for strongly agreeing. The comprised 10 questions consist of five positive and five negative statements. The SUS score is calculated using the collected data, and the result falls between 0 and 100. SUS is widely used in the literature to compare and evaluate the success of outputs.

The sense of presence experienced via exposure to an immersive environment has frequently been connected to the effectiveness of the environment. Presence is frequently described as a "feeling of being there" when a person feels as though they are in a different place. The Presence Questionnaire was first developed by Witmer and Singer (1998). It comprises several subsections, but only a few of them are incorporated in this study. Presence was rated on a 7-point Likert scale ranging from 1 = strongly disagree to 7 = strongly agree.

A methodology called the Technology Acceptance Model (TAM) seeks to determine whether new technology will be accepted, rejected, or usable. The approach was first presented by Davis (1987), and it is widely used in the literature. TAM consists of different subsections and different adoptions. A revised form of the TAM questionnaire (Venkatesh & Davis, 2000) is used and it consists of 10 questions that cover two subsections: perceived usefulness (PU) and perceived ease of use (PEU). The first six questions are used to measure PU, and the other four are used to calculate PEU. TAM was rated on a 7-point Likert scale ranging from 1 = strongly disagree to 7 = strongly agree.

The results of the effectiveness of the immersive systems, however, may be mediated by individual differences, aspects of the technology, and negative side effects. Measuring these mediating effects is of great importance in understanding the relationship between and among them, and how to maximize the effectiveness of the immersive environments. To measure several individual differences, the Immersive Tendency Questionnaire (Witmer and Singer, 1998) was introduced. The immersive tendency is an individual's proclivity to become immersed in the simulation. ITQ consists of 18 items and is composed of four subsections that include involvement (five items), focus (five items), emotions (five items), and propensity to play video games (three items).

To measure the tool competence level of participants regarding VR and MR, they were asked to rate their experience level on a 7-point Likert scale ranging from 1 = I have no experience to 7 = I am an expert. In the PC version, we utilized the most common navigation and interaction methods (W, A, S, and D buttons for navigation, and mouse interaction for camera movements), which are not required for MR and VR. Therefore, instead of asking about experience level in PC, we employed a survey called the Quantic Foundry Gamer Motivation Profile to measure participants' degrees of video game experience in the study. The questionnaire comprises six questions that focus on respondents' favorite games and game-playing habits and were created utilizing factor analysis and historical investigation. It assesses experience based on four levels, ranging from non-gamer to hard-core gamer.

Four additional open-ended question sets were answered by the users for this study, namely, Framework Competence, Interaction Preference, Procedural Generation Algorithm Selection, and Lighting Preferences. Tool Competence Questionnaire includes two questions to assess experience level regarding VR and MR. There is one question for each version (VR and MR) with two options (hand interaction/controllers) for interaction mode preferences. In the Lighting Preferences Questionnaire, there are two questions for different roles (curator/visitor) per version (VR and PC) with three temperature options (cold/neutral/warm). Procedural Generation Algorithm Selection consists of four questions for each version (based on data, based on user preferences) for three procedural generation algorithms to identify ease of use, perceived control, understandability of the algorithm, and the quality of the design outcomes. It is rated on a 5-point Likert scale ranging from 1 = none to 5 = completely.

## 5. Results

The comparative results presented in this chapter are based on the responses given by the 30 participants for the PC version with MacBook M1 Pro and VR and MR version with Oculus Quest 2. Responses to open-ended questions were analyzed using the analysis tool MAXQDA (2022). Statistical analyses were conducted using the software program JASP (2022 V.0.16.4) to compute and validate the results. For each questionnaire, reliability tests were run to determine whether the mean values' internal consistency was satisfactory. We compute Cronbach's alpha values, which is a method often used to validate surveys (Table 1).

Table 1. Interpretation of Cronbach alpha (α) values (Tavakol et al., 2001).

| Cronbach's α | Cronbach's α Comments |
| --- | --- |
| α ≥ 0.9 | Outstanding |
| 0.7 ≤ α < 0.9 | Good |
| 0.6 ≤ α < 0.7 | Acceptable |
| 0.5 ≤ α < 0.6 | Weak |
| α ≤ 0.4 | Unacceptable |

### 5.1. Semi-structured Interview

The first and second interview questions require users to explain their general impressions regarding the framework and their interpretation of the usage area. Figures 10, 11, and 12 present the frequency mapping of words that are commonly used to describe different versions of the framework and potential use cases.

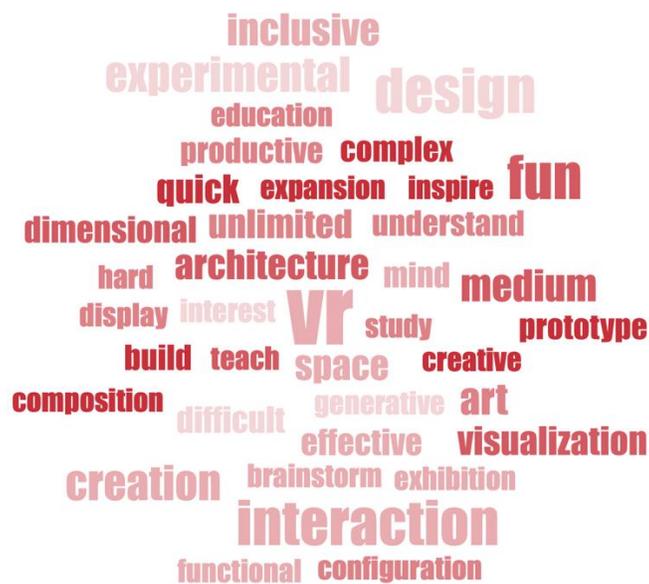

Figure 10. Frequency Mapping of Most Used Words by Participants to Describe the VR version of the framework. Produced with MAXQDA.

Figure 11. Frequency Mapping of Most Used Words by Participants to Describe the MR version of the framework. Produced with MAXQDA.

Figure 12. Frequency Mapping of Most Used Words by Participants to Describe the PC version of the framework. Produced with MAXQDA.

### 5.2. Interaction Preferences

Users were asked to select the medium that they preferred according to identified asset types. Figure 13 shows the users' preferences for interaction mediums with different asset types.

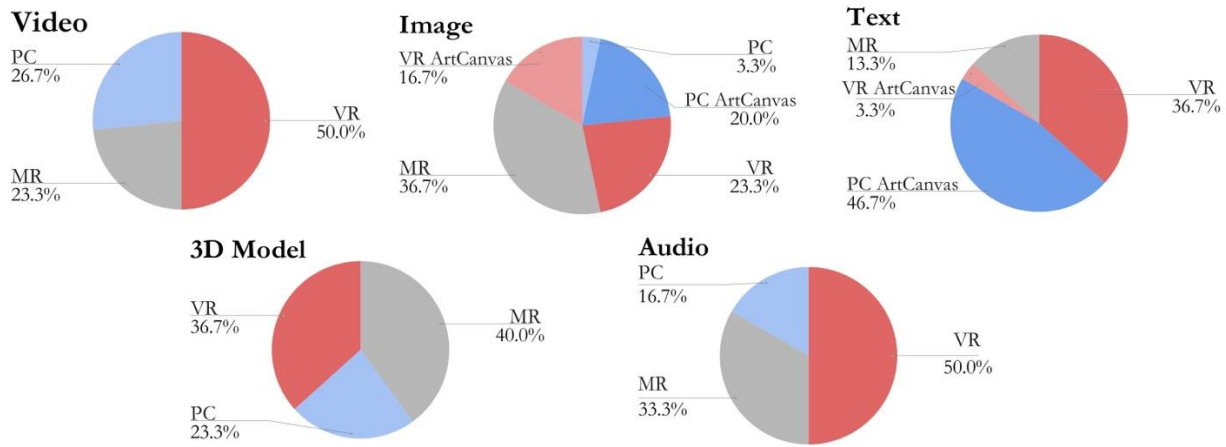

Figure 13. Interaction preferences of participants with different asset types.

In the VR and MR versions, all users tested the hand interaction and controllers. Figure 14 shows the users' preferences for interaction with different versions of the application.

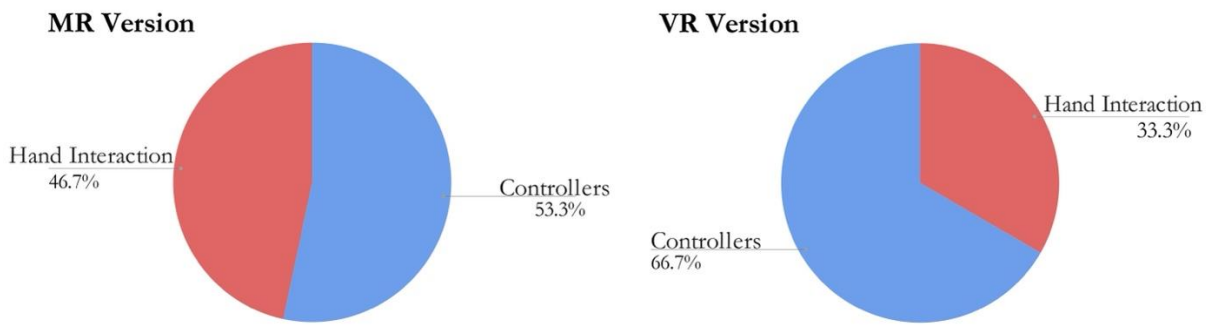

Figure 14. Interaction preferences of participants for VR and MR Version.

### 5.3. Lighting Temperature Preferences

Three parts of the temperature scale were given as a preference for the respondents for different modes of the framework. Since the MR version does not utilize the virtual lighting system, it was not included in the survey. The results are demonstrated in Figure 15.

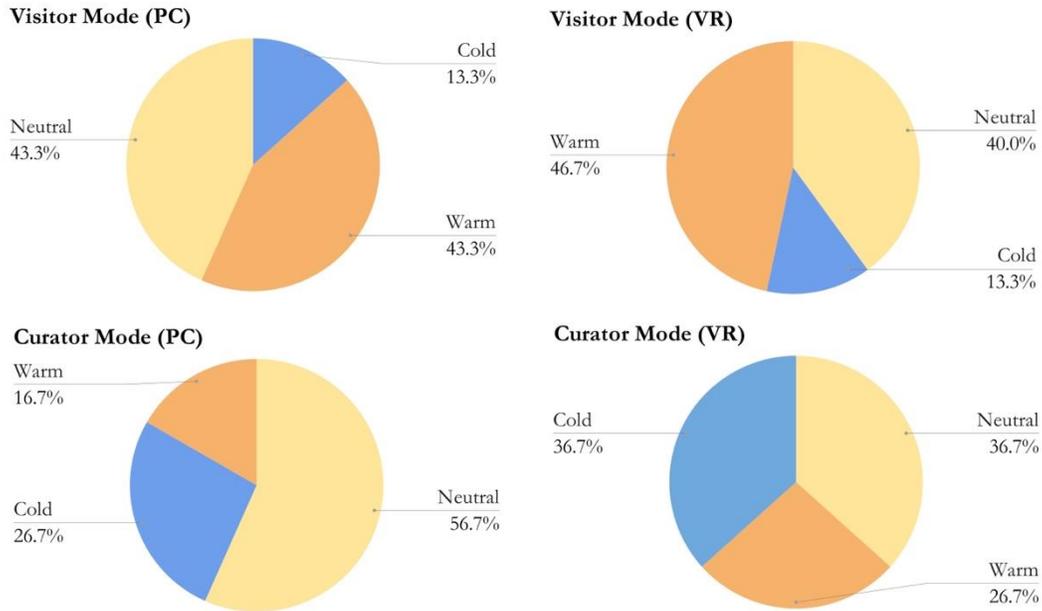

Figure 15. Lighting preferences of participants for different modes and immersive environments.

## 5.4. Procedural Generation Algorithms

Four different criteria were identified to assess the success of the implemented algorithms. The results are divided into two—generation based on data and generation based on user preferences. To test the correlation between the number of options provided to users and other eligibility indicators, they are presented together in Table 2.

Table 2. Mean and standard deviation results of the questionnaire for procedural generation based on user preferences. Answers to the questionnaires are on a 1 to 5 scale; a higher score indicates a more positive attitude.

|  | Growth Algorithm | Binary Space Partitioning and Cellular Automata | Room Generation |
|---|---|---|---|
| Number of Options | 3 | 7 | 4 |
| Perceived Control | 3.56 ± 1.04 | 3.76 ± 0.97 | 4.03 ± 0.80 |
| Ease of Use | 3.90 ± 0.75 | 3.7 ± 0.98 | 4.00 ± 0.78 |
| Comprehensibility | 4.00 ± 0.81 | 3.66 ± 0.92 | 4.00 ± 0.87 |
| Design Outcome | 3.54 ± 0.97 | 3.43 ± 0.89 | 4.03 ± 0.92 |

Since perceived control, the number of options, and ease of use are not relevant to generation based on data, only design outcome and comprehensibility of the algorithms were investigated. The results are given in Table 3.

Table 3. Mean and standard deviation results of the questionnaire for procedural generation based on data. Answers to the questionnaires are on a 1 to 5 scale; a higher score indicates a more positive attitude.

|  | Growth Algorithm | Binary Space Partitioning and Cellular Automata | Room Generation |
|---|---|---|---|
| Design Outcome | 3.23 ± 0.89 | 3.90 ± 0.75 | 3.40 ± 0.85 |
| Comprehensibility | 3.66 ± 0.71 | 2.86 ± 0.84 | 3.30 ± 0.83 |

The results show that although the comprehensibility of the combination of BSP and Cellular Automata algorithm is low in comparison to other algorithms, it has the highest score in terms of design outcome.

### 5.5. System Usability Scale

Although SUS does not have subscales, the answers should be computed in accordance with SUS scoring algorithms to interpret the results (Brooke, 1996). To create consistent scoring as required, the scoring formulas include different calculation methods for the questionnaire's negative and positive questions. Table 4 shows Cronbach's alpha values, while Table 5 and Figure 16 display the findings of SUS' statistical study.

Table 4. Cronbach's Alpha Values of System Usability Scale.

| Version | PC | VR | MR |
|---|---|---|---|
| Cronbach's $\alpha$ | 0.653 | 0.874 | 0.670 |

Table 5. Mean and Standard Deviation Results of System Usability Scale.

| Version | PC | VR | MR |
|---|---|---|---|
| Mean | 79.91 ± 10.89 | 70.00 ± 14.03 | 77.51 ± 12.99 |

The results of the SUS were found reliable since the calculated Cronbach's alpha ($\alpha$) value was higher than 0.5 for all versions. (PC (0.653 > 0.5), VR (0.874 > 0.5), MR (0.670 > 0.5)). The overall System Usability Scores (SUS) for 30 participants indicated high rates of system usability, with an average score of 79.91% for PC which is defined as "good" and is graded as "B" in the system usability assessment. Comparing the standard deviations reveals greater consistency in the participants' PC scores. On the other hand, the usability score of the MR version (77.51%) is higher than the VR version (70.00%), but both are in the range of "B" and are defined as "good" as well.

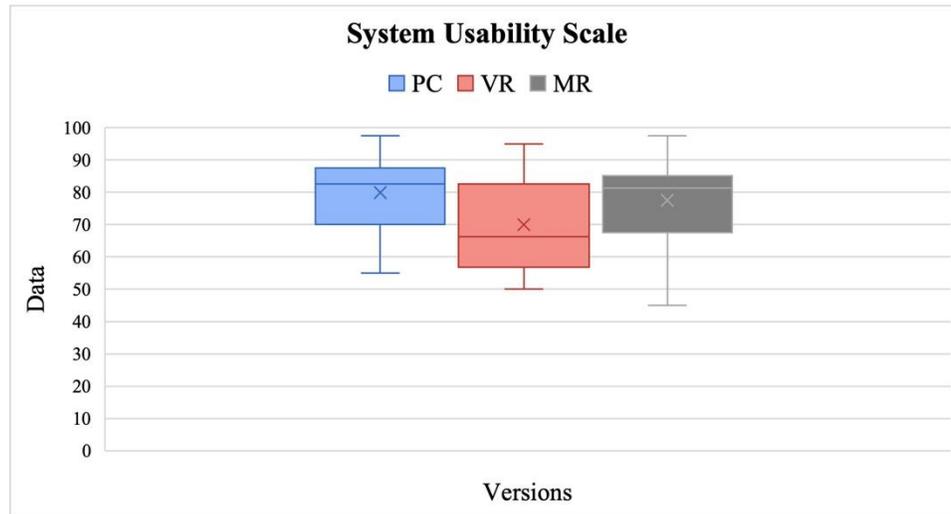

Figure 16. Comparative Results of SUS Scores in Boxplot Format.

### 5.6. Technology Acceptance Model

TAM questionnaire does not have a structured question set nor a scoring sheet to compare the results. Therefore, the results were examined only numerically using descriptive statistics. For the user study, the most related two subsections of the TAM model were selected, which are perceived ease of use and perceived usefulness. The questionnaire was analyzed with its subsection. The results are given in Table 6, Table 7, and Figure 17 comparatively for each version.

Table 6. Cronbach's Alpha Values of Technology Acceptance Model.

| Version | PC | VR | MR |
|---|---|---|---|
| Cronbach's α | 0.640 | 0.857 | 0.848 |

The collected data in total were found reliable since the calculated Cronbach's alpha (α) values were higher than 0.5 (PC (0.640 > 0.5), VR (0.857 > 0.5), MR (0.848 > 0.5)).

Table 7. Mean and standard deviation results of the Technology Acceptance Model. Answers to the questionnaires are on a 1 to 7 scale; a higher score indicates a more positive attitude.

| Version | PC | VR | MR |
|---|---|---|---|
| Perceived Usefulness | 5.63 ± 1.05 | 5.47 ± 1.22 | 5.5 ± 1.24 |
| Ease of Use | 5.54 ± 1.05 | 4.88 ±1.13 | 5.15 ± 1.41 |
| Total Mean | 5.60 ± 1.05 | 5.23 ± 1.22 | 5.29 ± 1.35 |

According to the results, PC's mean value, PU, and PUE scores were found to be higher than the VR and MR versions. While the MR version's score is slightly higher than the VR version, the most notable score is the PEU score of the VR version, which is lower than other versions.

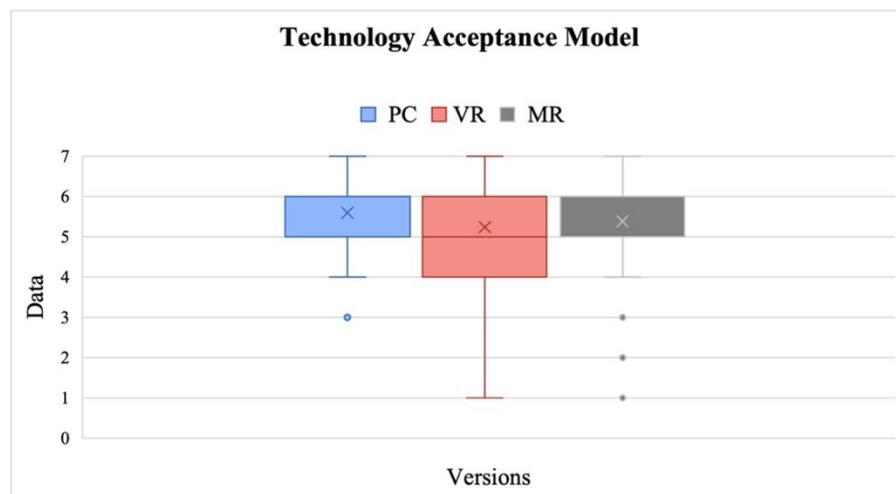

Figure 17. Comparative Results of Technology Acceptance Model Questionnaire in Boxplot Format.

## 5.7. Presence Questionnaire

The Presence Questionnaire (PQ) has subsections that represent diverse characteristics of the framework. By focusing on the subsections of the questionnaire, such as realism, possibility to act, quality of the interface, possibility to investigate, and self-evaluation of performance, a total of 19 questions were posed to assess versions' affordance in presence. Each subsection's overall system performance is calculated independently, and the results are then combined to get a total presence score. The scoring sheet provided by Laboratoire de Cyberpsychologie de l'UQO (L'UQO, 2002), which shows the minimum scores for a successful outcome, was used as a guideline for comprehending the presence capabilities of the successful versions. The results are given in Table 8, Table 9, and Figure 18 comparatively for each version.

Table 8. Cronbach's Alpha Values of Presence Questionnaire.

| Version | PC | VR | MR |
|---|---|---|---|
| Cronbach's α | 0.717 | 0.822 | 0.685 |

Table 9. Mean and standard deviation results of PQ. Answers to the questionnaires are on a 1 to 7 scale; a higher score indicates a more positive attitude.

| Version | PC | VR | MR |
|---|---|---|---|
| Realism | 5.26 ± 1.20 | 5.02 ±1.31 | 5.02 ±1.45 |
| Possibility to Act | 5.40 ± 1.15 | 5.25 ± 1.46 | 4.93 ± 1.37 |
| Quality of Interface | 4.37 ± 1.52 | 4.33 ± 1.56 | 4.93 ± 1.49 |
| Self-evaluation of Performance | 5.53 ± 0.92 | 5.38 ± 1.09 | 4.78 ± 1.43 |
| Total Mean | 5.14 ± 1.27 | 4.97 ± 1.46 | 4.93 ± 1.44 |

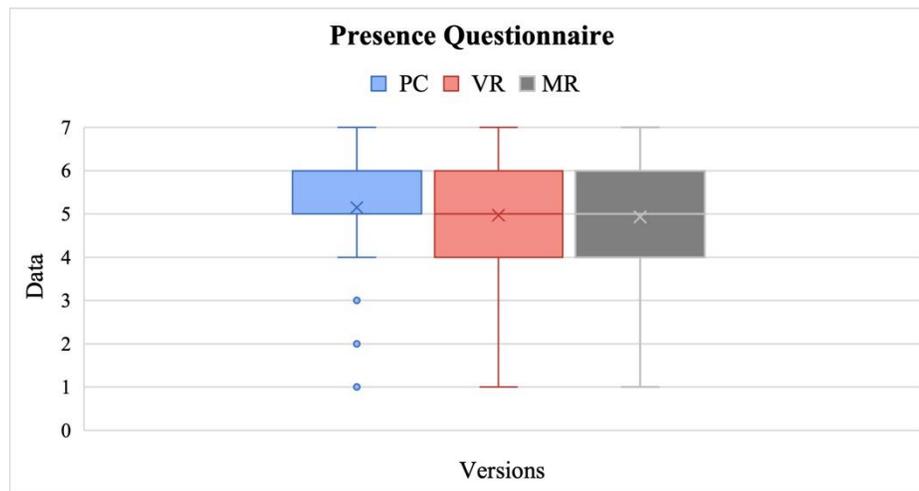

Figure 18. Comparative results of Presence Questionnaire in Boxplot Format.

According to the results of reliability tests, the results of the questionnaire were found eligible. While the total mean scores of versions are similar to each other, several outliers can be seen in Figure 18 for the PC version.

## 6. Discussion

Heterotopias are unique spaces that can connect different systems through the exchange of information and are characterized by the presence of multiple layers. Based on the features of discursive heterotopias, the framework described in this study is designed to be flexible and modular, which can be used in a variety of contexts to create digital habitats that can provide a space for discourse to occur. Because of the complexities of heterotopias, a variety of evaluation approaches are required to test the convenience of the methods regarding the objectives.

To test the framework with individual archives, while some participants prepared a collection of assets right before the sessions, some preferred to use existing materials in their online or offline databases. While the pre-made dataset provided us with essential insights into more technical aspects, the potential of the framework to construct unique heterotopias revealed itself with personal experiences and archives, proving the framework's applicability to diverse contexts.

The colocation of multimedia elements through the "operating table" approach has created several possibilities for participants. For example, participants preferred to use their photographs, videos, notes, and screenshots, similar to the accumulative aspect of heterotopias in terms of time and elements. They generated different rooms constructing "grids of specification" according to their categorization, which created multilayered experiences directly related to self-reflection and self-formation. Individual archives and spatial layers provided by the framework allowed users to create a "world within worlds."

[P.17] *"I considered the experience as walking inside my brain and organizing it, so it could work. Also, the framework can be extended via deep learning algorithms for further customization and guidance."*

Participants who used the assets related to their professional life have created relations and experiences based on "surfaces of emergence" that cannot be provided without place and embodiment. For example, participants from backgrounds related to the design studies used the models they produced together with the sketches, notes, and videos of production processes and presented narrative exhibitions. They created in-situ demonstrations and performed architectural design activities experiencing the creative process of "aesthetic becoming."

[P.22] *"I think most of the design tools that we used are like trucks; this application was like a sports car."*

[P.04] *"I can easily use this tool for my interior design practices, and I especially found the mixed reality version very useful."*

Participants with medical backgrounds placed anatomical models and texts, transforming the environment into a puzzle game where they could manipulate the parts of the model with hand interaction. Additionally, two participants from computer sciences added new features with dynamic models via scripts embedded in 3D models based on OOP using the adaptability of the framework. Throughout the semi-structured interview, we investigated the reasons behind the results of the questionnaires, participants' personal opinions about the versions of the framework, and their perception of the framework for possible use cases. Through the open-ended questions, we explored the interpretation of the framework from the users' perspective. Open-ended questions were used to gather data on how users perceive and interpret the framework. An analysis of the responses to these questions, using frequency mapping, showed that the different technologies used in the framework resulted in various interpretations. For example, words such as "game," "creative," "fun," and "art" were commonly used to describe the PC version, while the VR version was more frequently described using words such as "experimental," "design," "fun," "architecture," "visualization," and "inclusive." On the other hand, the MR version was often described using words such as "interior-design," "prototype," "interaction," and "transformation." A comparison of the frequency mappings suggests that MR and VR technologies may provide more comprehensive and multilayered experiences that involve embodied cognition. Also, most participants provided individual cases based on their experiences and needs.

[P.02] *"I would prefer to use this tool when I teach my students geography and history."*

[P.25] *"We can use this tool for designing, displaying, and teaching the systems in our mechanical engineering studies."*

The richness of the cases provided by the participants implies that the framework is applicable to various contexts. The dynamic autonomy layers and understanding of self-formation based on heterotopias provided design space for the users. They became the "focal points of resistance," producing informal use cases which were not structured by the framework but by the users.

The concept of heterotopia can be interpreted as a complex set of relationships which in this case, also involves the user experience which is affected by various elements from hardware to navigation. According to statements of respondents, different technologies and interaction modalities that are provided serve better for certain discourses. Also, according to the activities, the users change their preferences in environmental settings such as lighting and scale.

The multimedia elements of the discursive space were provided with different representation methods. Users' environment preferences differ according to media type. For the 3D models, users preferred to interact with them in VR and MR environments. Participants stated that MR provides better scale perception in relation to surroundings, and both other versions were more impressive. Most users prefer to display videos and images on a human scale. However, due to distractions created by physical objects in the MR version, the majority of users preferred the VR version for videos since it required a long-term focus. For the audio, most of the users preferred the VR environment, finding the combination of a completely artificial environment with audio more engaging. For the text interactions, the majority of the users tend to read in the ArtCanvas view, which is the most traditional option. Therefore, the dominant media format of discourse can affect the preferences of users in terms of technology.

When creating a multimedia system, it is important to consider the nature of the interaction and the anticipated results in terms of learning, sensory and emotional engagement, and satisfaction. The hand interaction and controllers were essential for conducting activities through embodiment. The hand interaction has similarities with the command language, and the users need to remember the gesture to complete the operation. Users mostly preferred to use controllers rather than hand tracking for interactions. Using buttons on controllers was more accessible and easier than hand gestures, especially for novice users. The results showed that there was an increase in the number of participants who preferred controllers in the MR version. This result suggests that being able to receive information from physical locations increases the request for physical body engagements.

It is possible to consistently stimulate human emotions and actions in virtual settings, which adds to the overall effectiveness of a heterotopia. The stimuli's visual integrity results in a high level of immersion. In this study, we looked into the lighting preferences of participants in the curator and visitor modes/roles. Results showed that there are differences in lighting preferences performing two different interactions. While the majority of the users preferred natural and cold light during the curation process, warm lighting settings were preferred during the visiting mode. In the VR environment, there is an increase in the selection of cold temperatures. The relationship between physiological input and an abstract conception in the curation process emphasizes that spatial perception is ultimately experiential and lived. Users' perceptions of and integration with virtual elements may change as a result of spatial perception blended with an embodiment. Commonly, users indicated that they were able to focus on details under cold and neutral lighting, and warmer temperatures made them feel more comfortable during observation.

Control and power mechanisms are important elements for both heterotopias and architectural practices. Heterotopias redefine those relations and provide an "order of discourse" with different syntaxes, where the syntax is constructed via places. The integration of technology changes the dynamics and defines the borders of control and power for the user. An increase in the autonomy of technology has created discussions in architectural practices. Architectural design practices cannot be distinctively separated from their tools. As technology advanced, design and art practices have changed their systems from physical to digital space. Architectural drafting, representation, building, and most critically, architectural design, have all been impacted by modern technologies, which have changed the production and final output. The general tendency of the tools points out the resistance of technology instead of users.

By providing a flexible framework and various autonomy levels for automated generation algorithms, we aimed to understand the dynamics between the technology, the user, and the transformation or construction of spatial layers of heterotopias provided via digital mediums. With the level of manual design provided to users, the overall outcome depends on how procedural generation is guided, constrained, and changed by the user. Moreover, since procedural generation is inherently deterministic, the integration of manual design steps is essential for users to inject variety and creativity. We aim to understand the impact of the level of autonomy on the outcome and user preferences. Here, we are particularly interested in the extent to which the method led to the intended results and higher levels of control, considering the tension between usability and comprehensibility associated with the design process of the users. Therefore, it is worthwhile to explore how the different outcomes are perceived by users.

According to the results, the growth algorithm provides the highest comprehensibility among the data-based versions. During the process, users were able to see the initial area that they or the algorithm would work on, which increased their understanding of how the algorithm works. In the user-based version, users can select the starting point for growth, but the size and shape of the rooms are determined by the algorithm, which the user cannot control. In the BSP version, the size of the rooms is chosen by the user, but their placement is controlled by the algorithm. The room generation algorithm allows the user to control both the placement and size of the final product. Overall, the growth algorithm provides a higher level of perceived control than BSP, but a lower level than the room generation algorithm. According to survey results, BSP combined with CA produces the best design outcomes but has a lower level of comprehensibility compared to the other algorithms. When participants generated the content based on their preferences, options that were given to users increased comprehensibility; however, this approach decreased the design outcome produced by the algorithm. This implies that the data-based version meets the space needs of the users better than the MR-based version. With the higher number of options presented, BSP provided a less perceived sense of control and ease of use. The data-based version of the room generation algorithm has lower scores for design outcome than BSP, but higher scores for perceived control, ease of use, and design outcome in the user-based version.

Most of the participants exhibit different strategies according to versions depending on their embodiment level, point of view, and scale. The spatial experiences in virtual environments provide an active interpretation process based on the awareness of surroundings. Therefore, the design approaches of participants differ according to versions. Comparing produced scales of the algorithms, BSP has the capability of producing large-scale designs while room generation can produce only one space, and the growth algorithm is able to produce within the borders of the footprint. Therefore, users tend to use the BSP algorithm in the PC version where they can observe the outcome with a completely top-down view. Room generation was mostly preferred in the MR version where users can change the surroundings with their virtual versions. The growth algorithm was found more suitable for the VR version since the users can produce multiple rooms, but still be able to navigate themselves without losing their tracks. Results also indicate that the comprehensibility of the method increases the perceived level of control.

Another objective of the study was to develop a framework for immersive technology based on heterotopias that is accessible and easy to use for a general audience, including both novice and experienced users where they can conduct discursive practices. The level of usability and acceptance can change an individual's control over the system by providing more resistance to the technology. Therefore, the framework was tested using the Technology Acceptance Model (TAM) and the System Usability Scale (SUS). The results showed that the different versions of the framework were convenient for users with different backgrounds and levels of competence, pointing out that the layers of heterotopia in a digital medium can provide an "operation table" for various cases and individuals. The SUS was used to assess the complexity, consistency, and ease of use of the system, and all three versions scored in a similar range. However, some participants preferred the VR version although it had lower usability scores, indicating that engagement may be a factor in the success of immersive technologies.

*[P.17] "I can use the VR version for hours when I want to get away from reality."*

*[P.04] "MR version was pretty interesting, the PC version was easier to use, but VR was more fun."*

According to user feedback analyzed through frequency analysis, the MR and VR versions of the framework were perceived as having both positive and negative qualities. While they were described as "experimental", "useful", "quick", "functional", and "interesting", they were also considered "difficult",

"hard", and "complex" to use. In contrast, the PC version was described as "easy "to use and more "game"-like. The high volume of information available within the heterotopias created an experimental and creative space for users but also resulted in interfaces that may be overwhelming for those who are new to the system. Some users noted that the numerous features hindered usability and increased cognitive load, disrupting their ability to fully immerse themselves in the environment. The intensification of information in heterotopias both produced "experimental" and "creative" ground for the users and also expressed itself via interfaces, which created exhaustion for novice users. Some participants indicated that the number of features decreases usability and increases the cognitive load, which disturbs the feeling of presence.

*[P.25] "I think the number of features should be less. I felt there were so many features to use, which put me under stress."*

To interpret the results of the TAM questionnaire in more detail, we examined the subsections. Even though the PC version had the highest scores, the results of the versions in terms of PU were similar to each other. The virtual reality (VR) version had the lowest PEU score, while the mixed reality (MR) version was perceived as being more understandable and requiring less effort. This difference may be due to the nature of VR, which can disconnect users from the real world. Some users found the MR version safer than the VR version, while the VR version was considered more enjoyable to use. The HMD used in the study creates a virtual boundary around the user and displays the real world when in close proximity to a physical object. However, some novice users who were new to being fully immersed in a virtual world may have had lower trust levels towards the technology and preferred to use the teleportation option for navigation, reducing their interaction with the virtual environment. In terms of performance, all three versions were produced based on the same system; therefore, they can sustain the same predictability level. However, other factors, reliability, and utility affect each other. In the case of VR, lower reliability to technology produces lower perceived usefulness results.

*[P.03] "Since I was able to see the physical environment, I felt safer in MR than VR."*

*[P.19] "I became disoriented in the VR, and at one point, I felt constrained by the headset and wanted to remove it. The MR version was not as impressive as the VR version, but I would prefer to use the MR version."*

The results showed that PC had higher presence scores than VR and MR, and this difference was attributed to the different levels of competence and familiarity that users had with these different technologies. Analysis of various subcategories, such as realism, quality of the interface, and the possibility of examination, found that each technology had its own strengths and weaknesses in these areas. For example, the PC version had higher scores in terms of naturalness of interaction and the ability to survey the environment, since the keyboard and mouse interaction are more natural for most of the users, and the PC version provides several camera options to the survey environment. On the other hand, the VR version had higher scores in terms of visual aspects and consistency with real-world experiences. The MR version had lower scores due to problems that occurred with hand interaction and the distractions caused by the colocation of physical and virtual content.

The results suggest that different technologies can have different impacts on cognitive load and users' focus, and it is important to consider these factors when choosing a technology for a particular task. Overall, total scores indicate that a multi-layer approach to constructing heterotopias with the given features provided high scores in presence, usability, and acceptance for users with different experience levels.

## 7. Conclusion

The discursive space refers to the understanding and insights gained through the study and examination of specialized knowledge about the world. With the increasing use of digital technologies, multimedia has become the primary means of communication in the digital age and the language of discursive practices. Heterotopias are special, multi-faceted contexts that facilitate the exchange of information and link to other systems with "surfaces of emergence." They are unique in their ability to connect and integrate various perspectives and knowledge bases. In this study, we presented a method for creating a framework for virtual and mixed reality environments, as well as personal computers, that allows for the exchange of knowledge and information through the creation of multilayered digital spaces based on heterotopias. The framework aims to fill in the gaps left by the disappearance of certain layers in archival practices, particularly spatial layers, due to the increasing reliance on technology that excludes embodiment. We provided the concept of a virtual museum as an "operating table" for discursive methodologies, and instead of static settings, we offered dynamic, "liquid architectures" which find their forms around the individual archives composed of elements of multimedia.

The framework was tested with 30 participants and was found to be useful, easy to use, and adaptable to various professional and personal settings. The effectiveness of the framework and transformative impact of heterotopias are presented via the results of the qualitative and quantitative approaches. We provided various sorting and grouping algorithms, and procedural content generation algorithms such as Binary Space Partitioning, Cellular Automata, Growth Algorithm, and Procedural Room Generation which offer different levels of autonomy to the automated generation algorithms, giving users an opportunity for reflection, modification, and control over the design. Results indicate that place-production, transformation, and archival practices have a mutual relationship that is essential for "self-formation" activities. Integration of the layers and features discovered within heterotopias with digital technologies, which offer a spatial "aesthetic becoming" at several levels of reality, was found productive for providing discursive practices within digital habitats. Blending the meta-design and end-user development approaches with the layers of heterotopias we provided sites where the users become "focal points of resistance."

To provide design features and outcomes that can direct future work in the HCI community, we analyzed the various factors that can influence user experiences and preferences. The results of the questionnaires and interviews demonstrate that the three evaluated versions are different, each having its own characteristics, strengths, and weaknesses. Through a comparative study, we demonstrate how different reality levels might augment different abilities of users by designing and curating virtual museums of their activity. The level of the hybridity of the physical and digital spaces, interactions, the meaning and perception of the space, and the elements that are contained by the space can be re-interpreted by different users in VR, MR, and PC environments.

## 8. Future Directions

This framework targeted essential features to provide a system for diverse contexts. Therefore, we did not add detailed technical requirements for a specific practice. Based on the framework structure, modules can be added to serve different professions. According to the literature review, most architects and designers prefer to work on several types of representations with multiple screens. With the additional features, this framework can solve colocation problems of the design processes and provide more detailed production. To reorder the content, we used four variables based on museology which can be reinterpreted by users. However, for more complex structures, features for different reordering methodologies can be generated. One of the factors that can affect the environment's realism is audio assets that are not provided in accordance with physical distance. To enhance the realism of the environment and navigate through these environments in a more realistic and immersive way, sound localization can be provided. According to participants' statements, the number of provided features can become exhausting for some novice users. Therefore, features can be provided separately for different experience levels.

In this study, we examined individual experiences separately for each technology to compare and understand specific preferences and impacts. As a future study, providing a social layer and more transformative systems between PC and immersive technologies can increase the richness and relations provided via heterotopias created by the users and increase the usability of the framework. Future studies can include asymmetric and symmetric collaborative approaches including the social relations to practices.

## Disclosure statement

No potential conflict of interest was reported by the author(s).

## Data availability statement

The data that support the findings of this study are available from the corresponding author, [E.S.], upon reasonable request.